\documentclass[preprint,12pt]{elsarticle}
\biboptions{sort&compress} 
\usepackage[T1]{fontenc}
\usepackage[utf8]{inputenc}
\usepackage{amsmath,amssymb,mathtools,bm,amsthm}
\usepackage{booktabs,array,longtable,multirow,tabularx}
\usepackage{graphicx}
\usepackage{subcaption}
\usepackage{float}
\usepackage{placeins}
\usepackage{xcolor}
\usepackage[colorlinks=true,linkcolor=blue,citecolor=blue,urlcolor=blue]{hyperref}
\usepackage{enumitem}
\setlist{nosep}
\graphicspath{{figures/}}

\usepackage[
 letterpaper,
 left=1.35in,
 right=1.35in,
 top=1.0in,
 bottom=1.0in
]{geometry}


\newcommand{\ii}{\mathrm{i}}
\newcommand{\dd}{\mathrm{d}}
\newcommand{\eps}{\varepsilon}

\newcommand{\calF}{\mathcal{F}}
\newcommand{\calG}{\mathcal{G}}

\newcommand{\calL}{\mathcal{L}}

\newcommand{\bfn}{\mathbf{n}}

\newcommand{\bfq}{\mathbf{q}}

\newcommand\figref[1]{Fig.~\ref{#1}}
\newcommand\tableref[1]{Table~\ref{#1}}

\newcommand\sectref[1]{Section~\ref{#1}}

\newcommand{\bfr}  {\mathbf{r}}
\newcommand{\bfd}  {\mathbf{d}}

\newcommand{\flameN}{\mathsf{N}}

\begin{document}

\begin{frontmatter}
\title{Conforming and nonconforming Trefftz approximations for two-dimensional scalar electromagnetic problems}
\author{Igor Tsukerman}
\address{Department of Electrical and Computer Engineering,\\
The University of Akron, Akron, Ohio 44325--3904, USA}
\ead{igor@uakron.edu}
\date{August 2026}

\begin{abstract}
Trefftz functions satisfy the differential equation locally and exactly; quasi-Trefftz functions do so approximately to prescribed high order. This paper considers (quasi-)Trefftz approximations for two-dimensional scalar electromagnetic problems. Established discretizations include the Flexible Local Approximation MEthod (FLAME), Trefftz elements ($T$-elements), and Trefftz discontinuous Galerkin (Trefftz-DG) methods. New developments are gradient-enriched FLAME (GEFLAME), conforming Trefftz--FLAME finite elements (TFF), and a full-field Bloch-wavevector-vs-frequency solution with Schur--DtN reduction. Applications cover electrostatics, scattering, singular fields, and Bloch waves in periodic structures.

These methods make different compromises between conformity and flexibility. FLAME and GEFLAME incorporate Trefftz functions directly into local difference schemes; TFF uses elementwise FLAME schemes to lift polynomial traces into element interiors; $T$-elements match elementwise Trefftz spaces weakly to such traces; and Trefftz-DG couples broken Trefftz spaces through fluxes and penalties.

In the reported wave-scattering examples, GEFLAME gives field and gradient errors several orders of magnitude below those of the quadratic finite-element discretization at comparable algebraic cost. Localized singular-function enrichment removes the dominant reentrant-corner error. Conforming TFF and quasi-conforming $T$-elements admit standard finite-element assembly but expose polynomial edge traces as an accuracy bottleneck. For periodic media, the bilinear Trefftz-DG formulation gives an analytic polynomial wavenumber-vs-frequency eigenproblem without restricting the Bloch multiplier to the unit circle. Schur--DtN reduction yields compact boundary-response problems.
\end{abstract}

\end{frontmatter}

\section{Introduction}\label{sec:Intro}
%
\subsection{Trefftz approximations}\label{sec:Trefftz-approximations}
%
One way of classifying discretization methods for boundary-value problems is via conformity vs flexibility of approximation. Conformity means that the numerical solution belongs to a global space with the required continuity properties; for the two-dimensional scalar electrostatic and electromagnetic problems on a computational domain \(\Omega\), the natural conforming space is \(H^1(\Omega)\). Flexibility means that the local approximation functions need not be polynomials; instead, a promising choice is \textit{Trefftz functions} which, by definition, satisfy the governing equation locally. Examples include harmonic functions for electrostatics and exponential ones for diffusion-layer problems; plane, cylindrical, or spherical waves in acoustics and electromagnetics. There is strong theoretical and numerical evidence for the approximation power of Trefftz functions: for elliptic and static problems, see the partition-of-unity and operator-adapted work \cite{MelenkBabuska1996,BabuskaMelenk1997,Melenk1999} and harmonic-polynomial Trefftz analysis \cite{HiptmairMoiolaPerugiaSchwab2014}; for Helmholtz and wave problems, see \cite{MoiolaHiptmairPerugia2011,HiptmairMoiolaPerugia2011,HiptmairMoiolaPerugia2016}. Recent results \cite{ParolinHuybrechsMoiola2023,GalanteMoiolaParolin2025} demonstrate that Trefftz bases enhanced by evanescent plane waves may significantly improve stability of local approximations.
The general idea, originally put forward in the 1926 paper by Erich Trefftz \cite{Trefftz-1926}, can be described in modern language as replacing generic local finite-dimensional approximation spaces with operator-adapted ones.

Exact Trefftz functions can be derived in many constant-coefficient or piece\-wise-constant subdomains. When exact local solutions are unavailable, \textit{quasi}-Trefftz functions can be constructed to satisfy the governing equation and applicable interface conditions to a prescribed local order \cite{ImbertGerardMonk2017,ImbertGerard2021,ImbertGerardSylvand2021,ImbertGerardMoiolaStocker2023,ImbertGerardMoiolaPerinatiStocker2025,FontanaImbertGerard2025}. 
%
\subsection{Objectives of the paper}\label{sec:Objectives}
%
The main motivation for this paper is to demonstrate that the versatility and breadth of Trefftz approximations is greater than commonly assumed. Trefftz and quasi-Trefftz functions can be used for different problem types and completely different discretizations.

Yet, the paper is not a survey. Established methods such as   Trefftz-discontinuous Galerkin (Trefftz-DG) and ultraweak variational formulation (UWVF)  \cite{CessenatDespres1998,HuttunenMonkKaipio2002,HiptmairMoiolaPerugia2011,HiptmairMoiolaPerugia2016}, the generalized finite element method (GFEM) \cite{MelenkBabuska1996,BabuskaMelenk1997}, Trefftz elements ($T$-elements) \cite{SzeLiu2010Tri,SzeLiu2010Quad}, and the Flexible Local Approximation MEthod (FLAME) \cite{Tsukerman2006,Tsukerman2010}, are presented only in condensed form to provide the relevant context.

\noindent
Specific contributions of this paper are as follows.
\begin{enumerate}
\item Gradient-enriched FLAME (GEFLAME) on structured and unstructured meshes. In addition to the field or potential values, gradients are included as degrees of freedom (DoFs), putting the field and its derivatives on an equal footing and aiming at comparable orders of accuracy for them. GEFLAME is introduced in \sectref{sec:unstructured-GEFLAME}; numerical tests are presented for single-cylinder scattering in \sectref{sec:Scattering-single-cylinder}, for four-cylinder scattering in \sectref{sec:four-cylinder-geflame}, and for alternative DoF layouts and meshes in Sections~\ref{sec:geflame-dof-layout-comparison} and~\ref{sec:geflame-cartesian-conforming-comparison}.
\item Conforming quasi-Trefftz-FLAME finite elements (TFF), in which polynomial skeleton traces are lifted into element interiors by local FLAME solutions on simple auxiliary grids. The construction applies to different element shapes and to elements cut by material interfaces. In the latter case, interface-matched Trefftz functions represent the local solution behavior without requiring a geometrically conforming material mesh. The construction is developed in \sectref{sec:FLAME-lift-construction}; the mixed-polygon and interface-cut electrostatic examples are presented in Sections~\ref{sec:Composite-polygon-electrostatic-FLAME-test} and~\ref{sec:Electrostatic-cylinder-FLAME-interface-cut}, and the Helmholtz comparison in \sectref{sec:Scattering-single-cylinder}.
\item Trefftz--FLAME basis enrichment in the presence of field singularities. Local enrichment is well established in FEM and especially in GFEM \cite{MelenkBabuska1996,BabuskaMelenk1997,BabuskaBanerjee2012}, but its use in FLAME has been limited \cite{AlKhateeb13}. The localized singular-function enrichment and the reentrant-corner numerical example are presented in \sectref{sec:Reentrant-corner-test}.
\item Application of Trefftz approximations to Bloch waves on periodic two-dimensional lattices (\sectref{sec:Bloch-tests}). Particular attention is paid to the \(q(\omega)\) eigenproblem, where \(q\) is the Bloch wavenumber and \(\omega\) is the frequency, rather than to the more traditional \(\omega(q)\) problem, and to the difference between the full-field and Bloch-periodic-factor formulations.
\item A full-field $q(\omega)$ Bloch construction in which Schur complements reduce assembled interior unknowns to a discrete Dirichlet-to-Neumann (DtN) boundary map while preserving polynomial dependence on the Bloch multiplier.  An earlier Bloch-band DtN approach was based on cylindrical-wave expansions and collocation \cite{Yuan-DtN-BandGaps-2006,Yuan-DtN-2007,Yuan-PhC-inteqn-DtN-2008}. Related finite-element Schur--DtN work \cite{KlindworthSchmidtFliss2014} is discussed in \sectref{sec:conforming-Schur-DtN}.
\end{enumerate}
The paper deals exclusively with two-dimensional electromagnetic problems. Extensions to 3D, relatively straightforward for some components of the methods and less so for the others, are envisioned as future work; see \sectref{sec:Conclusions}.
%
\subsection{Flexibility vs conformity}\label{sec:Flexibility-vs-conformity}
%
There is a persistent tension between flexibility of approximation and conformity. Classical polynomial elements provide global $H^1$ conformity but restrict the local approximation space, whereas Trefftz spaces are operator-adapted but cannot generally be matched strongly across elements. Nonconforming difference schemes such as FLAME can use Trefftz approximations directly \cite{Tsukerman2006,TsukermanCajko2008,Tsukerman2010}. TFF, $T$-elements, and Trefftz-DG make different compromises: TFF lifts a chosen conforming trace into a quasi-Trefftz interior; $T$-elements weakly match an interior Trefftz expansion to a shared trace; and Trefftz-DG retains broken Trefftz fields while controlling skeleton jumps and fluxes.
\subsection{Structure of the paper}\label{sec:Paper-structure}
The remainder of the paper is organized as follows. FLAME is recalled in \sectref{sec:FLAME}. Conforming TFF trace lifts are developed in \sectref{sec:FLAME-lift-construction}: the Laplace construction is presented in \sectref{sec:Rectangular-Laplace-element}, the curved-interface Helmholtz version in \sectref{sec:cut-interface-elements}, and the error structure in \sectref{sec:TFF-errors}. GEFLAME is introduced in \sectref{sec:unstructured-GEFLAME} and related to established methods in \sectref{sec:Existing-methods}. The numerical examples are organized by problem type: electrostatics in \sectref{sec:Electrostatic-examples}, Helmholtz scattering and singular-field enrichment in \sectref{sec:Helmholtz-examples}, and Bloch problems in \sectref{sec:Bloch-tests}.
%
\section{FLAME: Trefftz bases in difference equations}\label{sec:FLAME}
%
FLAME, the Flexible Local Approximation MEthod, constructs difference equations from problem-adapted bases rather than standard Taylor polynomials \cite{Tsukerman2006,Tsukerman2010}. Trefftz or quasi-Trefftz functions are most prominent in that context.

Let a grid molecule contain nodes \(\{\bfr_1, \ldots, \bfr_n\}\), and let \(\{\psi_1, \ldots, \psi_m\}\) be a set of basis functions satisfying the partial differential equation (PDE) and any applicable boundary conditions either exactly or approximately to high order near that molecule. Sampling these functions gives the local FLAME sampling matrix, denoted with \(\flameN\),
\begin{equation}\label{eqn:flame-sampling-matrix}
    \flameN \,=\, \bigl( \psi_j(\bfr_i) \bigr)_{i=1,\ldots,n; 
    \, j=1, \ldots, m}.
\end{equation}
A FLAME scheme is a nonzero vector \(s\in\mathbb C^n\) satisfying
\begin{equation}\label{eqn:flame-null-vector}
    s^\top \flameN \,=\, 0 \quad \Leftrightarrow \quad
    s \in \operatorname{Null}(\flameN^\top)
\end{equation}
so that the local equation \(\sum_i s_i u(\bfr_i)=0\) is exact on the selected basis. If $\operatorname{Null}(\flameN^\top)$ is one-dimensional, this determines the scheme up to normalization. 

When the nullspace is multidimensional, the nullspace condition \eqref{eqn:flame-null-vector} alone does not specify a unique scheme. A useful stability-oriented heuristic is to maximize diagonal dominance: for the equation associated with a designated DoF, choose the null vector whose target coefficient is largest relative to all remaining coefficients. Let us for convenience reorder the local DoFs so that this target DoF is first. Let $S$ be a matrix whose columns form an orthonormal basis of $\operatorname{Null}(\flameN^\top)$. Any linear combination $s = S c$, with a set of coefficients $c$, is then a potential FLAME scheme. Let us partition
\begin{equation}\label{eqn:flame-scheme-basis-partition}
    S=
    \begin{bmatrix}
        S_1\\
        S_2
    \end{bmatrix},
\end{equation}
where $S_1$ is the first row and $S_2$ is the remaining block. The selected scheme maximizes the target coefficient relative to all remaining coefficients, viz.
\begin{equation}\label{eqn:flame-multidimensional-scheme-selection}
    c_{\rm dd}
    \,=\,
    \underset{c \ne 0}{\operatorname{arg \, max}} \,
    \frac{|S_1 c|^2}{\|S_2 c\|_2^2},
    \quad
    S_1^*S_1 \, c_{\rm dd}
    \,=\,
    \lambda_{\max}S_2^*S_2 \, c_{\rm dd},
    \quad
    s_{\rm dd}
    \,=\,
    \frac{Sc_{\rm dd}}{\|Sc_{\rm dd}\|_2}.
\end{equation}
Thus the diagonal-dominance (``dd'') selection is a small generalized-eigenvalue problem; the final scalar normalization is done for convenience. This procedure can also be couched as the orthogonal projection onto the unit vector \( (1, 0, \ldots, 0 )^\top \).
In nodal-triplet GEFLAME, diagonal dominance optimization is applied separately to the field and the two derivative DoFs at the central vertex. 

If no exact null vector exists, for example because the basis is overcomplete, one instead applies SVD to minimize $\|s^\top \flameN\|_2$ subject to $\|s\|_2 = 1$.

The general idea of FLAME has several generalizations and various uses on regular and unstructured grids. Near a truncated exterior boundary, choosing outgoing waves as the local basis can produce local discrete radiation conditions \cite{TsukermanTrefftzMachine2014,PaganiniScarabosioHiptmairTsukerman2016}. In conforming TFF elements (\sectref{sec:FLAME-lift-construction}), FLAME schemes generate an elementwise lift from the boundary traces into element interiors. In GEFLAME (\sectref{sec:unstructured-GEFLAME}), the local set of DoFs is enriched with the field gradients, typically leading to a qualitative accuracy improvement. 
%
\section{Conforming TFF elements with local FLAME lifts}\label{sec:FLAME-lift-construction}
%
\subsection{A motivating example}\label{sec:Rectangular-Laplace-element}
%
The purpose of this section is to motivate the development of TFF elements; \sectref{sec:Electrostatic-examples} contains a full setup and numerical results.

\begin{figure}
\centering
\captionsetup{skip=3pt}
\captionsetup[subfigure]{skip=1pt}
\begin{minipage}{0.56\linewidth}
\centering
\begin{subfigure}{0.48\linewidth}
 \centering
 \includegraphics[width=\linewidth]{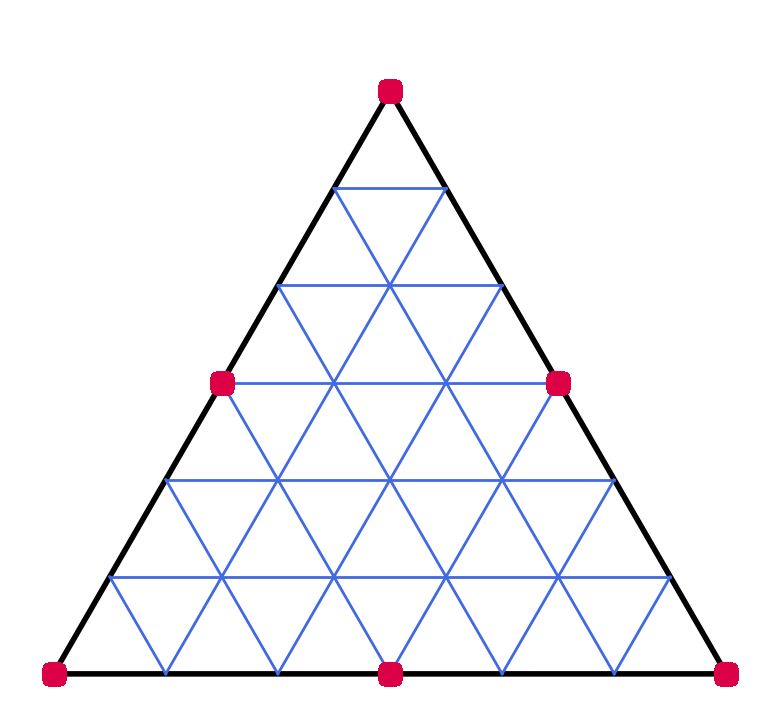}
 \caption{TFF trace lift on an equilateral triangle.}
\end{subfigure}\hfill
\begin{subfigure}{0.48\linewidth}
 \centering
 \includegraphics[width=\linewidth]{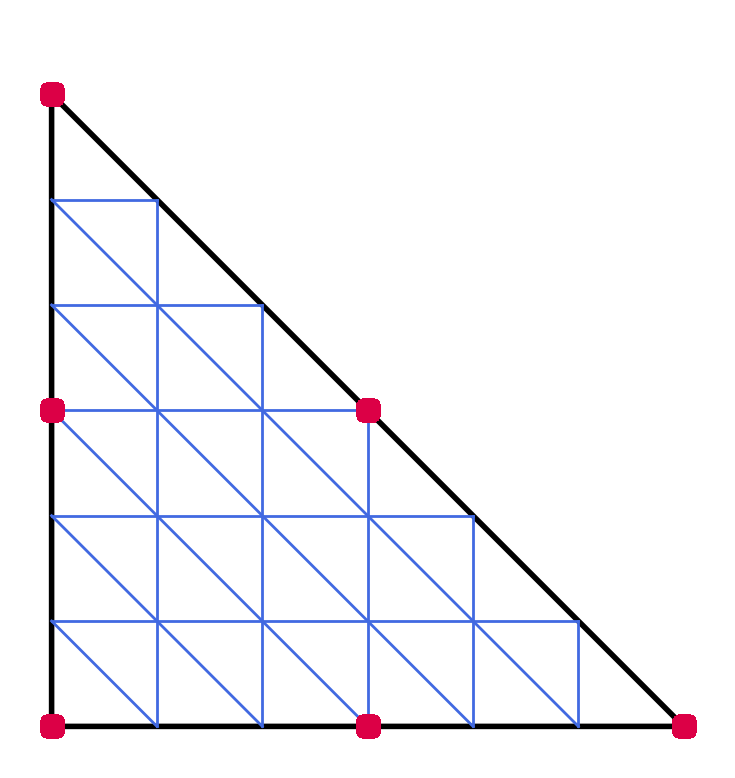}
 \caption{TFF trace lift on a right triangle.}
\end{subfigure}
\begin{subfigure}{0.48\linewidth}
 \centering
 \includegraphics[width=\linewidth]{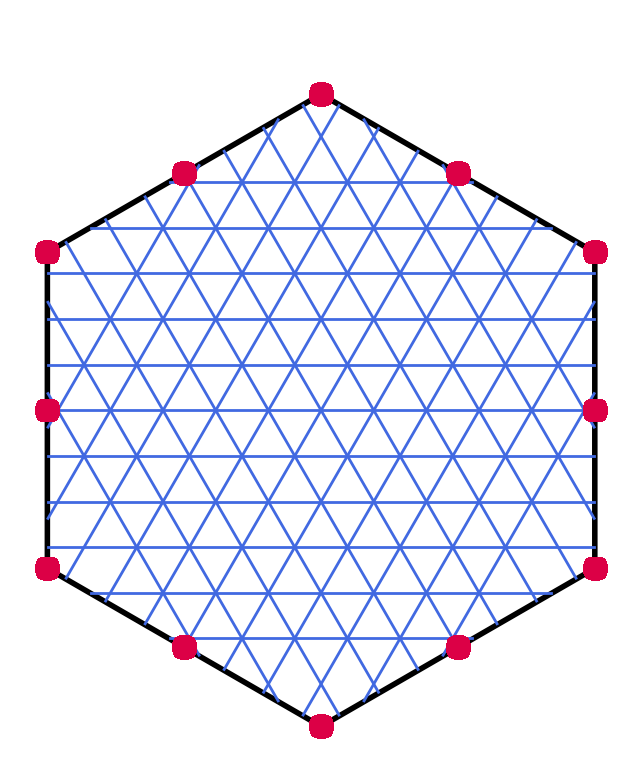}
 \caption{TFF trace lift on a regular hexagon.}
\end{subfigure}\hfill
\begin{subfigure}{0.48\linewidth}
 \centering
 \includegraphics[width=\linewidth]{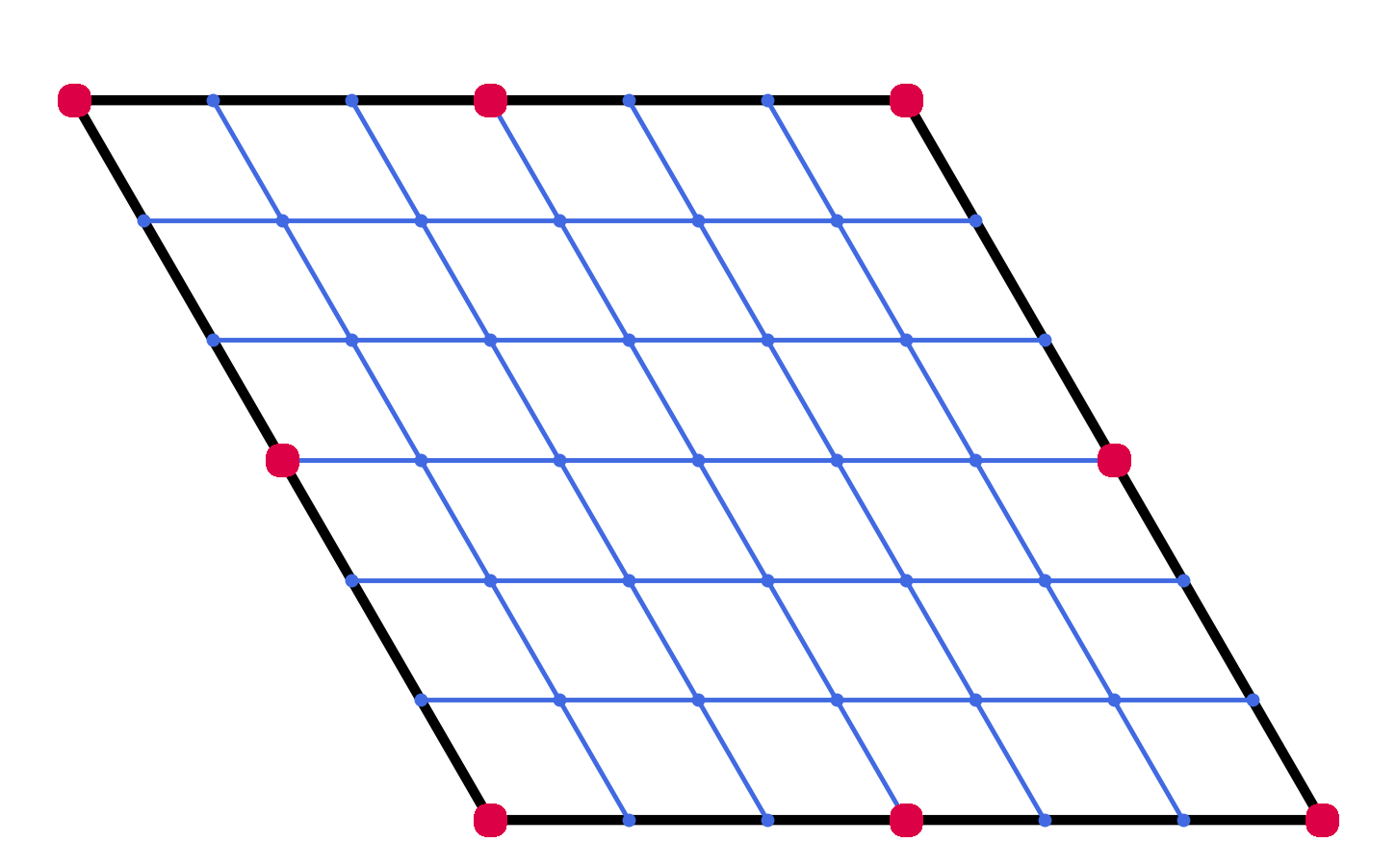}
 \caption{TFF trace lift on a rhombus.}
\end{subfigure}
\end{minipage}
\caption{TFF trace-lift construction: representative element-local auxiliary grids. Red points are the coarse $P_2$ trace nodes; blue lines show the subdivision used internally by the FLAME lift and for element-matrix construction.}
\label{fig:flame-grids-main}
\end{figure}

Consider a polygonal mesh, which may contain, in addition to triangles, rhombi and hexagons; unorthodox element shapes are introduced here deliberately to illustrate the versatility of the approach. The global DoFs are still the ordinary nodal values. Quasi-Trefftz basis functions within any given element are approximate solutions of the elementwise Dirichlet problem via high-order FLAME schemes on auxiliary FLAME grids, such as the ones shown in \figref{fig:flame-grids-main}. 

As an example, consider the electrostatic potential \(u\) satisfying the equation\footnote{Naturally, this example may be reinterpreted as heat conduction.}
\begin{equation}\label{eqn:tff-electrostatic-equation}
    \nabla \cdot (\eps \nabla u) \,=\, 0 .
\end{equation}
where the permittivity \( \eps \) is in general coordinate-dependent.
When \(\eps\) is constant inside an element \(K\), the local equation reduces to \(\nabla^2 u=0\). 

A quadratic Dirichlet trace on the element boundary is used in the simulations of \sectref{sec:Electrostatic-examples} as a meaningful example, but the approach is valid for any chosen trace. Each edge contributes its two endpoints and one midpoint DoF. Let \(t_K\) be the vector of these values corresponding to the quadratic edge traces.

Construction of the FLAME scheme on the auxiliary grid is standard; see summary in the previous section and \cite{Tsukerman2006}. In the reported Laplace implementation, the local Trefftz approximation set consists of the harmonic polynomials
\[
1,\ \operatorname{Re}z,\ \operatorname{Im}z,\ldots,\operatorname{Re}z^6,\ \operatorname{Im}z^6.
\]
The molecule row is selected by the minimum-singular-vector construction described in Section~2; when an exact left null vector exists, this reduces to the classical FLAME row.

Denote the resulting numerical FLAME lift from the edge traces into the element interior by \( \calL^{\rm F}_K t_K \).
This lift (i.e., the approximate Dirichlet solution) is then used to form the element stiffness matrix. To minimize the loss of accuracy in the spatial derivatives, gradients are computed by native FLAME interpolation -- that is, interpolation by the analytically computed gradients of the same Trefftz basis functions as were used to construct the FLAME scheme in the first place \cite{DaiTsukerman2008,Dai-Webb11,DaiThesis2009}. If \(D_K\) denotes the respective gradient-recovery operator at the chosen quadrature points and \(W_K\) the corresponding quadrature-weight matrix, the stiffness matrix contribution is
\begin{equation}\label{eqn:flame-lift-matrix-main}
  A_K^{\rm TFF} ~\approx~ \eps_K (D_K \calL^{\rm F}_K)^* 
  \, W_K(D_K \calL^{\rm F}_K).
\end{equation}
Here $^*$ denotes the conjugate transpose; in the real electrostatic examples it reduces to the ordinary transpose.
The global unknowns are still confined to the skeleton of the mesh, but each element uses a quasi-Trefftz approximation internally.
%
\subsection{Example: TFF-FLAME lifts for wave scattering from curved interfaces}\label{sec:cut-interface-elements}
%
A similar FLAME lift is applied in Helmholtz problems. Trefftz basis functions in 2D are typically taken as plane waves in homogeneous regions and as matched Bessel/Hankel harmonics in the vicinity of a circular interface. The algebra of interface matching conditions is straightforward \cite{Tsukerman2006}, \cite[\S 4.4.11, 8.10.6, 8.10.7, 8.12.5]{TsukermanBook2026} and is not reproduced here. Pieces of general smooth interface boundaries within a given grid molecule can be approximated by either arcs of the osculating circles \cite{AlKhateeb13} or, somewhat better, by optimal circular-arc fits (either way, a computationally inexpensive task as compared to the solver).

In homogeneous-medium elements, the local FLAME lift is based on harmonic polynomials for Laplace problems and on propagating, cylindrical, or evanescent waves for Helmholtz problems. For variable coefficients or local interfaces, FLAME lifts are derived from quasi-Trefftz approximations. TFF elements are conforming by construction. 

A standard FEM lift could in principle be used on a suitable local geometrically conforming mesh, but this auxiliary mesh generation is overkill in most practical cases. In contrast, a FLAME lift builds the local behavior directly into the Trefftz basis over simple grid molecules. In this approach, FLAME acts as an auxiliary tool for constructing elementwise quasi-Trefftz approximations, while the overall finite-element Galerkin framework remains in place. When the continuous problem is Hermitian, the same sesquilinear assembly preserves Hermiticity at the discrete level.
%
\subsection{Error decomposition and limiting mechanisms for TFF lifts}\label{sec:TFF-errors}
%
This section gives a qualitative error decomposition and identifies the principal factors limiting TFF accuracy. Sufficient solution regularity is assumed for the discussion to hold in the indicated Sobolev norms.

For one element $K$, let $\gamma_Ku:=u|_{\partial K}$ denote the exact
Dirichlet trace, and define the local degree-$p$ polynomial trace space
\begin{equation}\label{eqn:local-polynomial-trace-space}
M_p(\partial K)
~:=~
\bigl\{\lambda\in C^0(\partial K):
\lambda|_e\in\mathbb P_p(e)\ \text{for every edge }e\subset\partial K\bigr\}.
\end{equation}
Let $I_{p,K}\gamma_Ku\in M_p(\partial K)$ be the usual degree-$p$ nodal
interpolant of the exact trace on the element boundary. In a conforming mesh,
neighboring elements share the vertex and edge nodes and therefore receive
identical interpolated traces on their common edge.

Write $\calL_K$ for the exact elementwise Dirichlet lift and
$\calL_K^{\rm F}$ for the numerical FLAME lift. Since
$u=\calL_K\gamma_Ku$ locally and the lift is linear, the error splits as
\begin{equation}\label{eqn:tff-trace-lift-error-split}
  u-\calL_K^{\rm F} I_{p,K} \gamma_Ku
 ~=~
 \calL_K \bigl(\gamma_Ku-I_{p,K} \gamma_Ku \bigr)
  \,+\,
  \bigl( \calL_K-\calL_K^{\rm F} \bigr) I_{p,K} \gamma_Ku.
\end{equation}
The first term is the polynomial trace-interpolation error propagated into the
element interior by the exact lift. The second is the error of the numerical
FLAME lift applied to the interpolated trace.

\paragraph{Trace approximation}
Assuming that the exact
trace has regularity $H^\mu(\partial K)$, with
$1/2<\mu\le p+1$, the standard interpolation estimate has the form
\cite{Ciarlet1978,BrennerScott2008}
\begin{equation}\label{eqn:trace-approximation-hhalf}
  \| \gamma_Ku-I_{p,K}\gamma_Ku \|_{H^{1/2}(\partial K)}
   ~\le~
   C h_K^{\mu-1/2} \| \gamma_Ku \|_{H^\mu(\partial K)}.
\end{equation}
For a sufficiently smooth trace one may take $\mu=p+1$. Corners,
singularities, and intersections of element edges with material interfaces may
reduce the available value of $\mu$ and hence the observed convergence order.
The examples in this paper use $p=2$ unless otherwise stated.

\paragraph{Dirichlet lift stability}
For uniformly elliptic electrostatic elements, assume the standard stability
of the Dirichlet lift from $H^{1/2}(\partial K)$ to the interior energy norm
\cite{McLean2000,BrennerScott2008}. Then the first term in
\eqref{eqn:tff-trace-lift-error-split} satisfies
\begin{multline}\label{eqn:Dirichlet-trace-lift-bound}
  \| \calL_K(\gamma_Ku - I_{p,K} \gamma_Ku) \|_{H^1(K)}
  ~\le~
  C_{\rm lift}(K) \,
  \| \gamma_Ku - I_{p,K} \gamma_Ku\|_{H^{1/2}(\partial K)} \\
  \lesssim ~ 
  h_K^{\mu-1/2} \, \| \gamma_Ku \|_{H^\mu(\partial K)}.
\end{multline}
For a smooth trace, $\mu=p+1$, and the trace-lift contribution is of order
$h_K^{p+1/2}$.

For Helmholtz elements, let $k_K$ denote the local material wavenumber. The TFF
examples below are in the resolved regime $k_Kh_K\ll1$. Since the Dirichlet
eigenvalues on an element scale as $h_K^{-2}$, $k_K^2$ then lies well below the first local resonance, avoiding the coarse, pre-asymptotic regime in
which Helmholtz pollution becomes significant
\cite{Ihlenburg-Babuska-Helmholtz-I-1995,
Babuska-Sauter-pollution-1997,
Melenk-Sauter-wavenumber-explicit-2011,
HiptmairMoiolaPerugia2011,
HiptmairMoiolaPerugia2016}. 

\paragraph{FLAME lift accuracy}
Let $\calG_K$ be the set of  FLAME-grid nodes within a given element $K$ and write
\begin{equation}\label{eqn:flame-grid-discrete-norm}
\|v\|^2_{\ell^2(\calG_K)}
~=~
\sum_{\bfr_i\in\calG_K}|v(\bfr_i)|^2
\end{equation}
For a FLAME scheme of local consistency order $p_F$, assume the generic discrete estimate
\begin{equation}\label{eqn:flame-grid-error}
\|\calL_K\varphi-\calL_K^{\rm F}\varphi\|_{\ell^2(\calG_K)}
 ~\le~
C_{\rm F}h_K^{p_F}\|\varphi\|_{M_p(\partial K)} ,
\end{equation}
where $\|\cdot\|_{M_p(\partial K)}$ denotes the trace-space norm used in the
lift estimate.

Comparison with a continuous solution additionally requires an interpolation
from the auxiliary FLAME grid to the element interior; assume that the following estimate holds with some scheme-dependent and interpolation-dependent parameter $m$
\begin{equation}\label{eqn:flame-h1-stipulated}
  \| \calL_K \varphi-\calL_K^{\rm F} \varphi \|_{H^1(K)}
  ~\le~
  C_{\rm int} \, h_K^m \, \| \varphi \|_{M_p(\partial K)} ,
\end{equation}
In \sectref{sec:Electrostatic-cylinder-FLAME-interface-cut}, some elements
are cut by a material interface. The global trace remains polynomial, while
the local interior approximation uses interface-matched harmonics
\cite{Tsukerman2006,TsukermanCajko2008,TsukermanBook2026}. In addition to
trace interpolation and numerical lifting, the error may then contain
contributions from the finite interface-matched basis, the local circular-arc
representation of the interface, numerical quadrature, and linear algebra.
Collecting these contributions as $\delta_{\rm match}(K)$,
$\delta_{\rm arc}(K)$, and $\delta_{\rm quad}(K)$ gives the schematic local
estimate
\begin{multline}\label{eqn:combined-local-estimate}
  \|u-\calL_K^{\rm F} I_{p,K} \gamma_Ku \|_{H^1(K)}
  ~\lesssim~
  h_K^{\mu-1/2} \| \gamma_Ku \|_{H^\mu(\partial K)}
  ~+~ h_K^m \| I_{p,K} \gamma_Ku \|_{M_p(\partial K)}\\
 +~ \delta_{\rm match}(K) \,+\, \delta_{\rm arc}(K) \,+\, \delta_{\rm quad}(K).
\end{multline}
For a sufficiently smooth trace, the first term has order
$h_K^{p+1/2}$. If regularity is reduced by corners, singularities, or
interface crossings, $p+1$ is replaced by the available regularity index
$\mu$. This decomposition illustrates the error mechanisms involved.
%
\section{FLAME and GEFLAME on unstructured meshes}\label{sec:unstructured-GEFLAME}
%
Previous work on FLAME was mostly confined to regular rectangular grids, although more general setups were occasionally considered \cite{Tsukerman2010,Dai-Webb11}. It is therefore of interest to include here FLAME-related schemes on unstructured meshes. 

FLAME needs two ingredients: a set of DoFs and a good approximation space -- most prominently, spanned by Trefftz or quasi-Trefftz functions; there is a broad array of choices for both. The DoFs may be, e.g., nodal values, edge circulations or fluxes. Edge- and flux-based choices are natural in electromagnetic discretization, as in N\'ed\'elec/Whitney-type finite elements \cite{Nedelec1980,Nedelec1986,Bossavit1998}; FLAME, however, is a  nonconforming difference method \cite{Tsukerman2006,Tsukerman2010}. An arbitrary pairing of basis functions and DoFs cannot be expected to produce a good scheme, but many meaningful combinations exist.

A method introduced in this paper is \textit{gradient-enriched FLAME} (GEFLAME), whereby the nodal values of scalar-field gradients are added to the field itself as DoFs. While this increases the overall number of unknowns, one aims at a more-than-commensurate accuracy improvement. This enrichment has a direct physical motivation. Indeed, gradients of electromagnetic potentials represent physical fields; likewise, gradients of one-component fields in electrodynamics are closely related to the companion field, which deserves to be approximated with comparable accuracy. In the GEFLAME setting, one may view grid nodes as triplets, each carrying three DoFs -- the scalar field and its spatial derivatives. 

For the Helmholtz $s$-mode (one-component electric field) examples below, the triplet is normalized as $(E,k_0^{-1}\partial_x E,k_0^{-1}\partial_y E)$, making all three components dimensionally compatible. For unit permeability, the last two components are, up to a $90^\circ$ rotation, the transverse magnetic field. Since multiplication of every gradient component by the same nonzero constant does not affect relative errors, these errors, denoted below by $e_{\nabla E}$, are identical for $\nabla E$ and $k_0^{-1}\nabla E$.

To evaluate the role of Trefftz approximations, it is instructive to build FLAME and GEFLAME using the same underlying geometric mesh as for, e.g., second-order ($P_2$) triangular elements. (One GEFLAME graph node represents a triplet of field and gradient values.) Numerical experiments are reported in \sectref{sec:unstructured-flame-cylinder}.
%
\section{Relations to established methods}\label{sec:Existing-methods}
%
\subsection{Partition-of-unity GFEM}\label{sec:GFEM-partition-of-unity}
%
The generalized finite element method (GFEM) uses partition of unity to seamlessly merge disparate and otherwise nonconforming finite-element spaces, including notably Trefftz spaces \cite{BabuskaIhlenburgPaikSauter1995,MelenkBabuska1996,BabuskaMelenk1997}. However, partition of unity adds a significant layer of complexity, leading to linearly dependent approximating functions in some cases, to cumbersome numerical quadratures in the Galerkin method \cite{PlaksTsukermanFriedmanYellen2003}, and to complications in Dirichlet boundary conditions. These features can complicate implementation and conditioning, the attractive combination of conformity and flexibility notwithstanding.
%
\subsection{Trefftz-DG, ultraweak and virtual element methods}\label{sec:DG-ultraweak}
%
Trefftz-DG and ultraweak variational formulations (UWVF) use local Trefftz functions and impose interelement coupling through numerical fluxes and penalties \cite{CessenatDespres1998,HuttunenMonkKaipio2002,HiptmairMoiolaPerugia2011,HiptmairMoiolaPerugia2013,HiptmairMoiolaPerugia2016,MoiolaPerugia2018}. The use of plane waves, cylindrical or spherical harmonics is backed up by extensive approximation theory and improves the accuracy for moderate-to-large wavenumbers. As a trade-off, careful jump control with adjustable penalty parameters is necessary.
The Trefftz virtual element method of \cite{MascottoPerugiaPichler2019} is more
abstract but more geometrically flexible. It defines local Trefftz or FLAME spaces
implicitly, through Helmholtz equations in the element interior and
plane-wave-generated impe\-dance traces on the boundary. 
%
\subsection{Quasi-Trefftz approximations}\label{sec:Quasi-Trefftz}
%
Quasi-Trefftz functions constructed in \cite{ImbertGerardMonk2017,ImbertGerard2021,ImbertGerardSylvand2021,ImbertGerardMoiolaStocker2023,ImbertGerardMoiolaPerinatiStocker2025,FontanaImbertGerard2025}
satisfy variable-coefficient partial differential equations (PDEs) to high local order when exact Trefftz functions are not easily available. One notable example is generalized plane waves for wave propagation in smoothly inhomogeneous media \cite{ImbertGerardMonk2017}: local phase and amplitude expansions lead to basis functions satisfying the variable-coefficient Helmholtz equation up to a prescribed order. Subsequent quasi-Trefftz DG work involved the accuracy, conditioning, and approximation properties of generalized plane waves  \cite{ImbertGerard2021,ImbertGerardSylvand2021,FontanaImbertGerard2025}, as well as extensions to space--time wave problems and smooth-coefficient elliptic problems \cite{ImbertGerardMoiolaStocker2023,ImbertGerardMoiolaPerinatiStocker2025}. In \cite{StockerVoulis2026}, quasi-Trefftz DG bases are obtained within a standard polynomial element space by enforcing the Helmholtz equation in a local weak sense.

Quasi-Trefftz functions have been extensively used in FLAME \cite{Tsukerman2006,Tsukerman2010}.
%
\subsection{A Trefftz continuous Galerkin method}\label{sec:Trefftz-continuous-Galerkin}
%
The Trefftz continuous Galerkin (TCG) method of \cite{GalanteDespresParolin2025} is conforming and locally-Trefftz for homogeneous two-dimensional Helmholtz problems. It constructs globally continuous, compactly supported basis functions on a rectangular mesh, using a combination of evanescent plane waves. This approach therefore departs from the usual discontinuous Trefftz-DG setting: the discrete space itself is continuous, and the element matrices can be assembled from Trefftz functions without volume residual terms. The formulation in \cite{GalanteDespresParolin2025} is developed for homogeneous two-dimensional Helmholtz problems on rectangular meshes and does not address variable coefficients, material interfaces, or general unstructured meshes.
%
\subsection{\texorpdfstring{$T$-elements}{T-elements}}\label{sec:T-elements-principles}
%
Trefftz finite elements have a long history in computational mechanics and numerical PDEs \cite{Herrera2000,JirousekZielinski1997}. The triangular and quadrilateral $T$-elements \cite{SzeLiu2010Tri,SzeLiu2010Quad}, included in this paper for comparison, use Trefftz spaces inside each element, weakly matched to given polynomial traces on the element boundary. The DoFs are, ultimately, the same nodal quantities as in standard polynomial FEM.

This is close in spirit, but opposite in direction, to the FLAME-lift construction of TFF elements (\sectref{sec:FLAME-lift-construction}). Namely, while $T$-elements start with exact elementwise Trefftz approximations and match those weakly to the prescribed boundary traces, the auxiliary elementwise FLAME schemes start with the traces and lift them to quasi-Trefftz functions in the interior.

One complication in $T$-elements is ill-conditioning, typical for many Trefftz bases. For the Helmholtz equation, several Trefftz basis functions (usually, plane waves) become nearly indistinguishable in their boundary traces, especially for wavelengths much longer than the element size, and the trace-fitting matrix can be extremely ill-conditioned. For FLAME lifts, conditioning is regularized by the auxiliary FLAME grid. Interestingly, recent publications \cite{ParolinHuybrechsMoiola2023,GalanteMoiolaParolin2025} have shown that  evanescent plane waves (EPWs) included in a Trefftz basis may lead to stable approximations, in contrast with bases containing only propagating plane waves. This suggests possible further enhancements for $T$-elements and other Trefftz methods as well.
%
\subsection{Static condensation and discontinuous enrichment}\label{sec:static-condensation-comparison}
%
Static condensation, in its classical finite-element form, eliminates internal degrees of freedom by a local Schur complement and leaves a reduced system on the retained interface variables \cite{Guyan1965,Irons1965,StrangFix1973}. If the FLAME lift described in \sectref{sec:FLAME-lift-construction} were replaced by a standard FEM on a submesh of each element, then the resulting method could indeed be classified as  static condensation.

Another relevant point of comparison is discontinuous-enrichment methods, whereby the standard finite-element polynomials are supplemented by discontinuous elementwise functions. Interelement compatibility is imposed weakly, via Lagrange multipliers or DG-type coupling \cite{FarhatHarariFranca2001,FarhatHarariHetmaniuk2003}. This, however, is conceptually closer to modern Trefftz-DG and ultraweak methods \cite{CessenatDespres1998,HuttunenMonkKaipio2002,HiptmairMoiolaPerugia2011,HiptmairMoiolaPerugia2016} than to TFF elements. 

\subsection{Summary of methods}\label{sec:Summary-of-methods}
%
In our discussion, it is instructive to classify Trefftz-related methods as nonconforming vs. weakly or strongly conforming (depending on the kind of trace matching between adjacent elements) and, separately, as Trefftz vs quasi-Trefftz. In principle, this gives several possible combinations, with different trade-offs between generality and accuracy. For example, one may hope for spectral convergence in conforming Trefftz methods, but they exist only in special circumstances \cite{GalanteDespresParolin2025}.
\tableref{table:method-taxonomy} compiles the terminology and characterization of various methods and elements in the remainder of the paper. 

\begin{table}
\centering
\caption{Terminology for method families used in the numerical sections.}
\label{table:method-taxonomy}
\footnotesize
\setlength{\tabcolsep}{2pt}
\renewcommand{\arraystretch}{1.10}
\begin{tabularx}{\textwidth}{|
>{\raggedright\arraybackslash}p{0.142\textwidth}|
>{\raggedright\arraybackslash}p{0.164\textwidth}|
>{\raggedright\arraybackslash}p{0.136\textwidth}|
>{\raggedright\arraybackslash}p{0.194\textwidth}|
>{\raggedright\arraybackslash}X|}
\hline
Family & Unknowns & Mesh & \shortstack{Global coupling /\\conformity} & Basis / lift \\
\hline
$P_2$ FEM & Nodal scalar field & Triangular & Shared conforming trace & Polynomial \\
\hline
FLAME & Nodal field or potential values & Regular or unstructured & Nonconforming & Trefftz or quasi-Trefftz \\
\hline
GEFLAME & Field and gradient values & Regular or unstructured & Nonconforming & Trefftz or quasi-Trefftz \\
\hline
TFF elements & Polynomial trace DoFs & Polygonal & Shared conforming trace & Quasi-Trefftz; FLAME lift \\
\hline
$T$-elements & Polynomial trace DoFs & Triangular or rectangular & Weak trace matching & Trefftz \\
\hline
TFF interface-cut & Polynomial trace DoFs & Interface-cut polygonal & Shared conforming trace & Interface-matched quasi-Trefftz; FLAME lift \\
\hline
Trefftz-DG & Element-wise Trefftz coefficients & Material-aligned cells & Flux/field penalty coupling & Trefftz \\
\hline
\end{tabularx}
\end{table}

\section{TFF elements in electrostatics}\label{sec:Electrostatic-examples}
%
\subsection{Formulation}
\label{sec:Electrostatic-problem-formulation}
%
The electrostatic potential $u$ satisfies
\begin{equation}\label{eqn:div-eps-grad-u-eq-0}
  -\nabla \cdot\bigl(\eps(\bfr) \nabla u(\bfr)\bigr) = 0
  \quad\text{in } \Omega,
  \quad
  u=g \quad \text{on } \partial \Omega .
\end{equation}
with the standard regularity assumptions for the boundary and Dirichlet values $g$. In the examples below, the coefficient $\eps$ is assumed piecewise constant, as the most typical case in practice. Inside each
constant-coefficient element $K$, \eqref{eqn:div-eps-grad-u-eq-0} reduces to the Laplace equation
\begin{equation}\label{eqn:Laplace-constant-coefficient-element}
    \nabla^2 u=0 .
\end{equation}
Thus harmonic polynomials are natural local Trefftz functions in this case. 
Across material interfaces, the potential and normal flux are continuous,
\begin{equation}\label{eqn:electrostatic-interface-conditions}
    [u]=0,\quad [\eps \, \partial_n u] \,=\, 0,
\end{equation}
with the usual \( [\, , \,] \) notation for the jump; $\partial_n$ implies a common normal direction in both subdomains.
Unless otherwise stated, the primary discrete metric is the relative root-mean-square (RMS) error $e_{\rm rel}$ over \(N\) discrete samples:
\begin{equation}\label{eqn:relative-rms-error-squared}
 e_{\rm rel}^2
 ~:=~
 \frac{N^{-1}\sum_{j=1}^{N}|v_h(x_j)-v_{\mathrm{cmp}}(x_j)|^2}
   {N^{-1}\sum_{j=1}^{N}|v_{\mathrm{cmp}}(x_j)|^2}
 ~=~
 \frac{\|v_h-v_{\mathrm{cmp}}\|_2^2}{\|v_{\mathrm{cmp}}\|_2^2}.
\end{equation}
Here \(v_h\) denotes the computed field and \(v_{\mathrm{cmp}}\) denotes a chosen comparison field: the exact or manufactured solution when available, or an independently converged comparison field otherwise.
The same convention is used for nodal vectors, vertex-grid samples, midline samples, and fine-grid postprocessing samples, as identified in the numerical results below.

The variational treatment of the continuous Dirichlet problem is standard. Choose any auxiliary function $g_0\in H^1(\Omega)$ whose trace on $\partial\Omega$ is $g$, and write $u=g_0+w$ with $w\in H_0^1(\Omega)$. With the bilinear form
$a(\phi,v):=\int_\Omega \eps\,\nabla\phi\cdot\nabla v\,\dd\bfr$, find $w\in H_0^1(\Omega)$ such that
\begin{equation}\label{eqn:electrostatic-continuous-weak}
    a(w,v)=-a(g_0,v)
    \quad \forall v\in H_0^1(\Omega).
\end{equation}
The electrostatic calculations below use the TFF elements of \sectref{sec:FLAME-lift-construction}, with quadratic traces and native FLAME recovery of the potential gradients. Once the local lifts are available, the element matrices and global conforming system are assembled as in standard FEM, and Dirichlet trace DoFs are eliminated in the usual manner.
%
\subsection{Composite-polygon electrostatic test for TFF elements}
\label{sec:Composite-polygon-electrostatic-FLAME-test}
\begin{figure}
\centering
\includegraphics[width=0.6\linewidth]{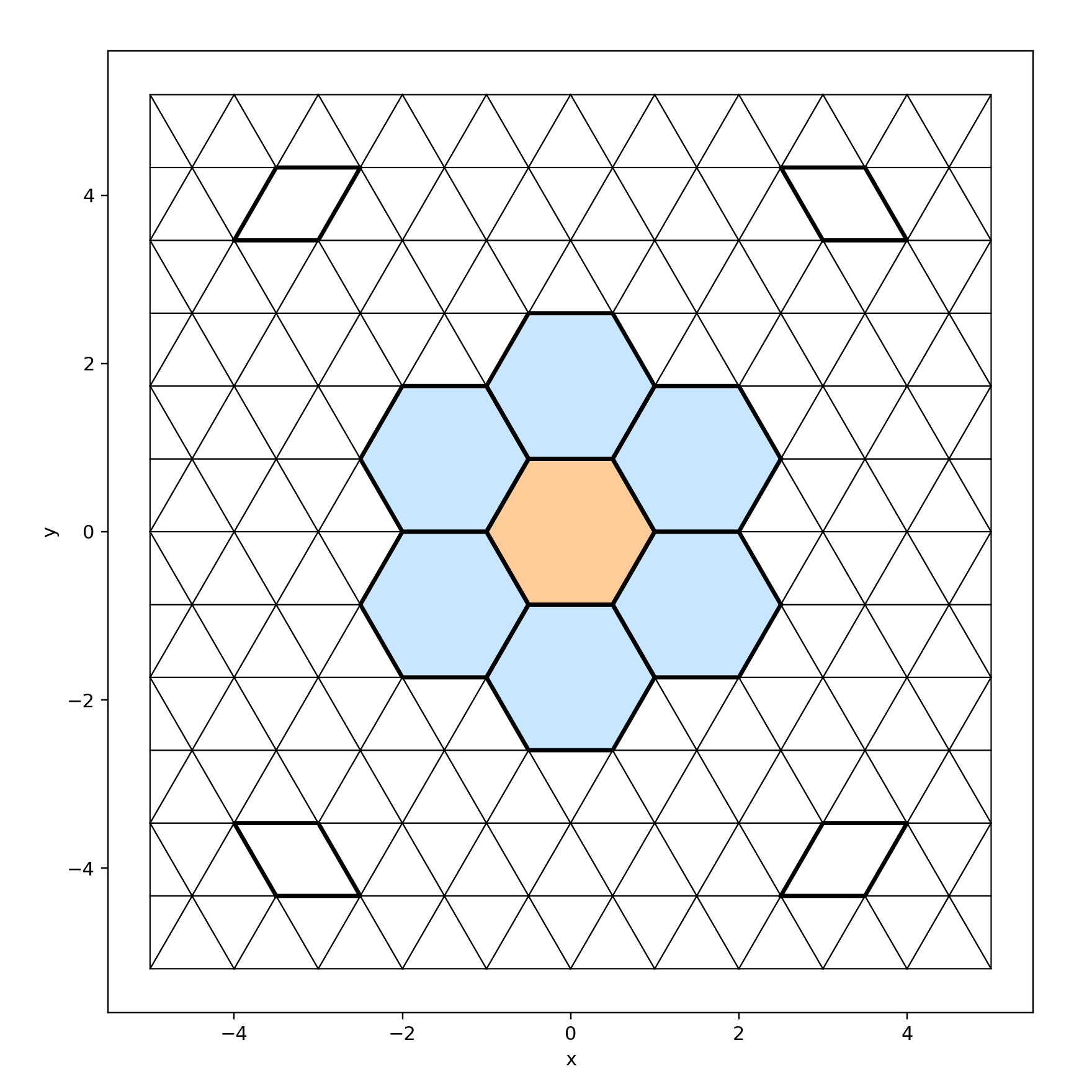}
\caption{Composite-polygon electrostatic TFF example: material mesh. The central hexagon has \(\eps=10\), the six surrounding hexagons have \(\eps=5\), and all remaining elements have \(\eps=1\).}
\label{fig:electrostatic-flame-material-mesh}
\end{figure}

%
The governing equation solved in this section is \eqref{eqn:div-eps-grad-u-eq-0}. To illustrate the versatility of TFF elements, this example deliberately uses an unorthodox mixed polygonal mesh shown in \figref{fig:electrostatic-flame-material-mesh}, with triangles, rhombi, and hexagons.

The standard quadratic trace on each element edge is lifted into the element's interior by an auxiliary FLAME scheme (\sectref{sec:Rectangular-Laplace-element}). More precisely, parametrize an edge \( [\bfr_1, \bfr_2] \) by $\bfr(t) = (1-t) \bfr_1 + t\bfr_2$, $0 \le t \le1$. The usual quadratic Lagrange functions on the edge are
\begin{equation}\label{eqn:quadratic-edge-Lagrange-functions}
    \ell_0(t) = (1-t)(1-2t),\quad
    \ell_1(t) = 4t(1-t),\quad
    \ell_2(t) = t(2t-1),
\end{equation}
$\ell_1$ being the edge bubble and $\ell_0,\ell_2$ the endpoint functions. For each \( \ell_{0,1,2} \), the lift into the element's interior is obtained by solving the local Dirichlet problem with that quadratic trace on the selected edge and zero trace on the remaining edges. This is accomplished by the degree-6 harmonic FLAME lift described in Section~3.1 on the auxiliary grids of \figref{fig:flame-grids-main}. Note that FLAME is used to construct elementwise quasi-Trefftz bases and does not contribute any global unknowns.

For the reported calculation, the central hexagon has $\eps=10$, the six surrounding ones have $\eps=5$, and all remaining elements have $\eps=1$. The particular Dirichlet conditions on the exterior boundary are not central to the illustration; for simplicity,
\begin{equation}\label{eqn:u-eq-x}
    u(x,y)=x
    \quad 
    \text{on } \partial\Omega .
\end{equation}

For comparison, an alternative simulation on a geometrically conforming mesh with second-order ($P_2$) equilateral triangles is also carried out. This mesh is obtained in a natural way by keeping the existing triangles, splitting each rhombus into two triangles, and subdividing each hexagon into six triangles by connecting its center to its vertices. The relevant mesh statistics are listed in \tableref{table:electrostatic-flame-mesh-dofs}.

\begin{table}
\centering
\caption{Composite polygon mesh and global skeleton sizes for the conforming TFF electrostatic test.}
\label{table:electrostatic-flame-mesh-dofs}
\begin{tabular}{|l|c|c|}
\hline
Quantity & TFF skeleton & $P_2$ comparison\\
\hline
Elements & 213 polygons & 252 triangles\\
\hline
Hexagons / triangles / rhombi & 7 / 202 / 4 & --\\
\hline
Global quadratic-trace DoFs & 496 & 549\\
\hline
Dirichlet DoFs & 88 & 88\\
\hline
Free DoFs & 408 & 461\\
\hline
Nonzero matrix entries & 5800 & 4413\\
\hline
\end{tabular}
\end{table}

\begin{figure}
\centering
\includegraphics[width=0.82\linewidth]{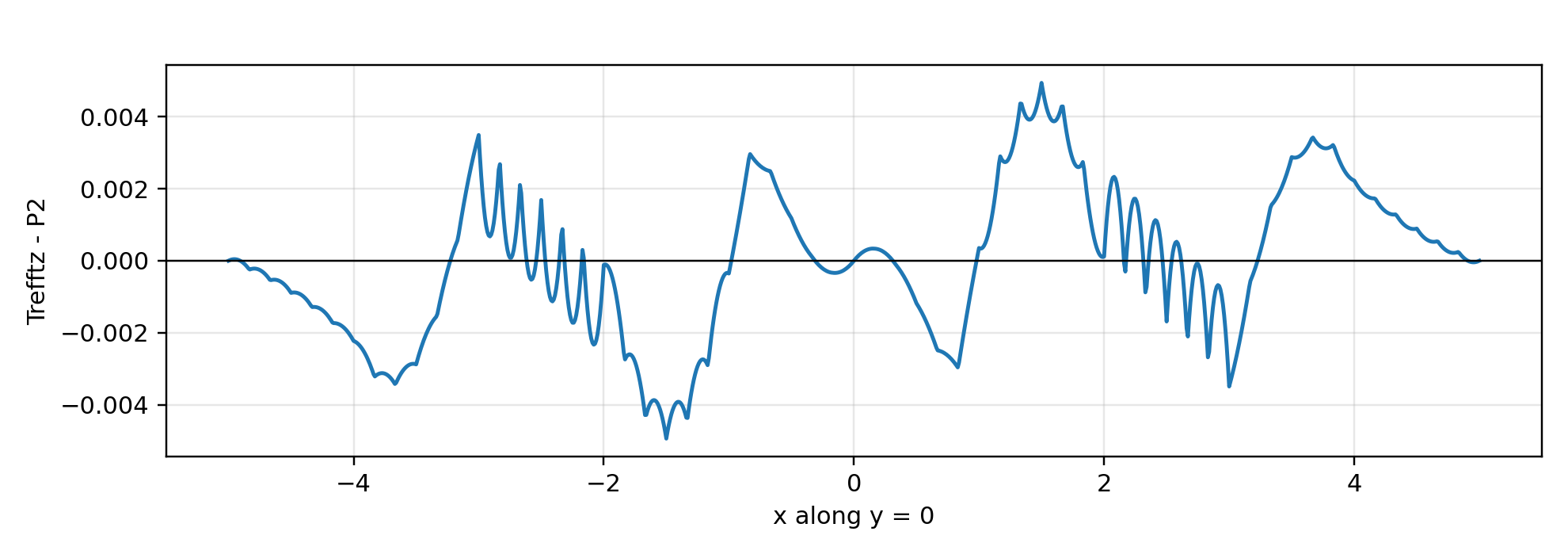}
\caption{Composite-polygon electrostatic TFF example: horizontal-midline difference between the TFF and comparison $P_2$ solutions.}
\label{fig:electrostatic-flame-midline-diff}
\end{figure}

\begin{figure}[H]
\centering
\includegraphics[width=0.85\linewidth]{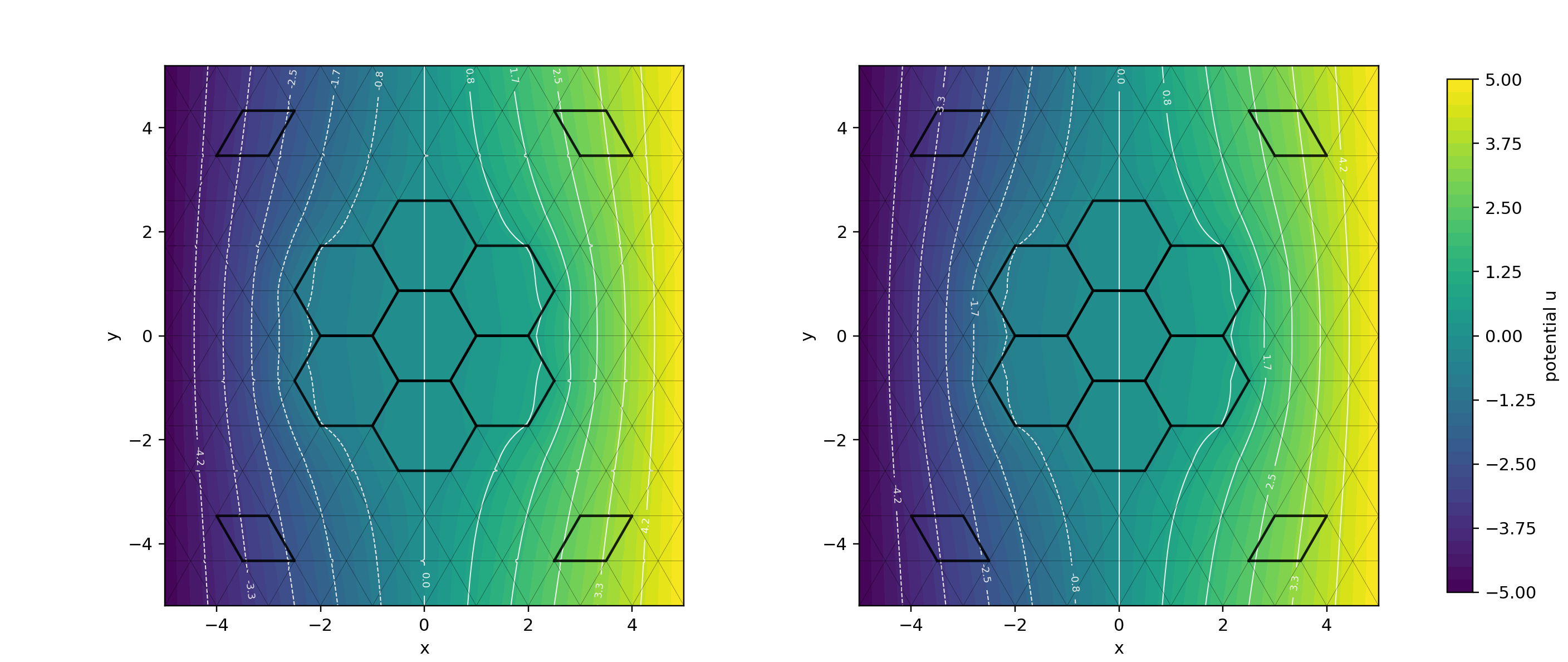}
\caption{Composite-polygon electrostatic TFF example: side-by-side contours of the TFF solution and the comparison $P_2$ triangular finite-element solution.}
\label{fig:electrostatic-flame-contour-comparison}
\end{figure}

The horizontal midline comparison used 1001 samples at $y=0$. Along that line, the maximum absolute difference between the conforming TFF solution and the $P_2$ comparison solution was $4.94\times10^{-3}$, and the root-mean-square (RMS) difference was $2.05\times10^{-3}$.
The  small difference along the horizontal midline is plotted in \figref{fig:electrostatic-flame-midline-diff}, and the full-field contour plots are shown in \figref{fig:electrostatic-flame-contour-comparison}. 

\FloatBarrier
\subsection{Rectangular TFF elements for a dielectric cylinder example}
\label{sec:Electrostatic-cylinder-FLAME-interface-cut}
%
The second TFF electrostatic test involves a dielectric cylinder on a regular rectangular grid. For error evaluation purposes, the Dirichlet data corresponds to the exact static interface solution for a uniform applied field in the \(y\)-direction,
\begin{align}\label{eqn:u-eq-r-sin-theta-etc}
 u_{\rm out}(r, \theta)
  & ~=~ -r\sin\theta
   \,+\, \frac{\eps_{\rm cyl}-\eps_{\rm host}}
       {\eps_{\rm cyl}+\eps_{\rm host}}
    \frac{r_{\rm cyl}^2}{r} \sin \theta,
    && r > r_{\rm cyl},\\
 u_{\rm in}(r,\theta)
  & ~=~ -\frac{2\eps_{\rm host}}
      {\eps_{\rm cyl}+\eps_{\rm host}} \,
    r \sin \theta,
    && r < r_{\rm cyl}.
\end{align}
The Dirichlet values $u_{\rm out}(r,\theta)$ from \eqref{eqn:u-eq-r-sin-theta-etc} are imposed on the outer square boundary, so \eqref{eqn:u-eq-r-sin-theta-etc} is also the exact solution used to evaluate the numerical errors.
 
An unconventional feature of this example is a regular rectangular grid, in which some rectangular elements cut through the circular interface; thus, the medium within some elements is not uniform. The DoFs, however, are standard nodal values associated with polynomial edge traces. The edge values are shared between neighboring elements, so the method is conforming. 

As explained earlier, the TFF basis functions are constructed via the FLAME solution of an elementwise Dirichlet problem on an auxiliary grid. FLAME schemes themselves are derived from harmonic polynomial bases in each medium, with the standard potential and flux matching at circular boundary \cite{Tsukerman2006}:
\begin{align}
  \psi_m^c(r, \theta)
  &=
  \begin{cases}
  \beta r^m \cos m \theta, & r \leq r_{\rm cyl},\\
  \left( r^m - \eta \, r_{\rm cyl}^{2m}r^{-m} \right) \cos m\theta, & r \geq r_{\rm cyl},
  \end{cases}\\
  \psi_m^s(r, \theta)
  &=
  \begin{cases}
  \beta r^m \sin m \theta, & r \leq r_{\rm cyl},\\
  \left( r^m - \eta \, r_{\rm cyl}^{2m}r^{-m} \right) \sin m \theta, & r \geq r_{\rm cyl}.
  \end{cases}
\end{align}
where 
\begin{equation}
   m = 1,2,\ldots; 
   \qquad
    \eta \,=\, \frac{\eps_{\rm cyl} - \eps_{\rm host}}
        {\eps_{\rm cyl} + \eps_{\rm host}},
    \qquad
    \beta \,=\, \frac{2\eps_{\rm host}}
          {\eps_{\rm cyl} + \eps_{\rm host}} .
\end{equation}
No logarithmic monopole is included in this interface basis. 

The electrostatic equation, the corresponding bilinear form and consequently the stiffness matrices involve gradients. Hence, the numerical accuracy of the latter is critical; yet standard finite difference approximations of derivatives are typically low-order. A more accurate interpolation of both the potential and its gradient can be effected via the native FLAME basis functions -- the ones used to construct the scheme in the first place. Within an interior FLAME-grid plaquette, the gradient is computed by analytically differentiating and averaging the local FLAME interpolants centered at the four surrounding FLAME-grid vertices. Near the element boundary, where not every vertex has a corresponding full \(3 \times 3\) molecule, the gradient is averaged over the available nearest complete molecules.

In the illustrative example, $r_{\rm cyl}=1$, $\eps_{\rm cyl}=4$, $\eps_{\rm host}=1$, and the outer domain is $[-4,4]^2$; these  parameters are also used later in the single-cylinder scattering case. The two meshes contain \(32 \times 32\) and \(64 \times 64\) rectangular elements. Because interface crossings reduce the regularity of the polynomial edge traces, straightforward trace $p$-refinement is not expected to be effective without resolving or enriching those crossings. Consequently, no attempt has been made to systematically study $p$-convergence; the edge trace order is limited to \(p=1\) and \(p=2\). An auxiliary \(20 \times 20 \) FLAME grid is used in each element. Relative RMS errors are reported in \tableref{table:electrostatic-cylinder-flame-results} for the non-Dirichlet nodal values, and, separately, for the sampled values along the vertical midline \( x = 0 \), where the exact solution has the largest variation.

\begin{table}
\centering
\caption{Relative RMS errors for rectangular TFF elements.}
\label{table:electrostatic-cylinder-flame-results}
\setlength{\tabcolsep}{4pt}
\begin{tabular*}{\textwidth}{@{\extracolsep{\fill}}|c|c|c|c|c|c|}
\hline
level & elements & \(p\) & DoFs & \shortstack{nodal rel.\\RMS} & \shortstack{midline rel.\\RMS}\\
\hline
0 & \(32\times32\) & 1 & 1089 & \(1.925\times10^{-3}\) & \(2.259\times10^{-3}\)\\
0 & \(32\times32\) & 2 & 3201 & \(5.665\times10^{-4}\) & \(5.842\times10^{-4}\)\\
1 & \(64\times64\) & 1 & 4225 & \(7.904\times10^{-4}\) & \(1.110\times10^{-3}\)\\
1 & \(64\times64\) & 2 & 12545 & \(1.917\times10^{-4}\) & \(2.235\times10^{-4}\)\\
\hline
\end{tabular*}
\end{table}

\FloatBarrier
\section{Numerical examples for Helmholtz problems}\label{sec:Helmholtz-examples}
%
\subsection{Model problem and error conventions}\label{sec:Helmholtz-problem-formulation}
%
Examples in this section involve, for concreteness, $s$-polarization, whereby a one-component electric field \( E \) satisfies the variable-coefficient Helmholtz equation in the Gaussian system of units,
\begin{equation}\label{eqn:Helmholtz-model}
    \nabla^2 E(\bfr) + k_0^2 \eps(\bfr) E(\bfr) \,=\, 0,
    \quad 
    k_0 \,=\, \frac{2\pi}{\lambda_0},
\end{equation}
with free-space wavenumber $k_0$, free-space wavelength $\lambda_0$, and problem-specific relative permittivity \(\eps(\bfr)\). For this polarization, both  field and its normal derivative are continuous across dielectric interfaces,
\begin{equation}
    [E] = 0, \qquad [\partial_n E] = 0 .
\end{equation}
where the square brackets indicate the jumps of \( E\) and of its derivative relative to a common normal direction for two adjacent regions.
For scattering examples, the total field is decomposed into incident and scattered components in the standard way:
\begin{equation}\label{eqn:E-eq-Einc-plus-Escat}
    E ~=~ E^{\rm inc} + E^{\rm scat}.
\end{equation}
Numerical operators \( \mathcal{R} \) approximating the radiation conditions for the scattered field \( E^{\rm scat} \) have the generic form 
\begin{equation}\label{eqn:R-Escat-approx-0}
  \mathcal{R} E_{\nu}^{\rm scat} ~\overset{\nu}{\to}~ 0
\end{equation}
asymptotically with respect to a set \( \nu \) of discretization and other parameters\footnote{For example, \( \nu \) may comprise the mesh size, the wavelength, separation between the scatterers and the exterior boundary, etc.}. Specific definitions of \( \mathcal{R} \) are given and used in the examples below. 

Since Trefftz basis functions are most naturally defined for the full field \( E \), one uses splitting \eqref{eqn:E-eq-Einc-plus-Escat} to translate \eqref{eqn:R-Escat-approx-0} to the full-field condition
\begin{equation}\label{eqn:R-E-approx-R-Einc}
  \mathcal{R} E_{\nu} ~\overset{\nu}{\to}~ \mathcal{R} E_{\nu}^{\rm inc}
\end{equation}
for the chosen \( \mathcal{R} \). Relative RMS errors reported in the examples below are defined analogously to those of \sectref{sec:Electrostatic-problem-formulation}.
%
\subsection{Scattering from a circular dielectric cylinder}\label{sec:Scattering-single-cylinder}
%
\paragraph{Formulation and analytical solution}\label{sec:Cylinder-scattering-exact-solution}
%
In this case, the relative permittivity is
\begin{equation}\label{eqn:cylinder-permittivity-Helmholtz}
 \eps(\bfr) \,=\,
 \begin{cases}
   \eps_{\text{cyl}}, & r < r_{\text{cyl}},\\
   \eps_{\text{host}}, & r > r_{\text{cyl}},
 \end{cases}
\end{equation}
where \((r,\theta)\) are polar coordinates and \(\theta=0\) is the positive \(x\)-axis. Define the host and cylinder wavenumbers by
\begin{equation}
  k_{\text{host}}=k_0\sqrt{\eps_{\text{host}}},
  \quad
  k_{\text{cyl}}=k_0\sqrt{\eps_{\text{cyl}}}.
\label{eqn:k-host-k-cyl}
\end{equation}
The incident plane wave with polar propagation angle \(\alpha\) is
\begin{equation}\label{eqn:incident-plane-wave-alpha}
  E^{\rm inc}_{\alpha}(r, \theta)
  \,=\,
  \exp \bigl(\ii k_{\text{host}} r\cos(\theta-\alpha)\bigr).
\end{equation}
The standard cylindrical-harmonic representation follows from the Jacobi--Anger expansion and separation of variables at a circular interface; see, for example, \cite{Twersky1952ParallelCylinders} and \cite[\S\S 8.10.6--8.10.7]{TsukermanBook2026}. For numerical evaluation of the comparison field, the scattered and interior series are truncated to orders \(|n|\le N_{\max}\).
\begin{equation}\label{eqn:E-exact-cyl-vs-rtheta-out}
  E_{\rm out}(r,\theta)
  ~=~
  E^{\rm inc}_{\alpha}(r,\theta)
  +
  \sum_{n=-N_{\max}}^{N_{\max}}
  a_nH_n^{(1)}(k_{\text{host}}r)\exp(\ii n\theta),
  \quad r > r_{\text{cyl}}.
\end{equation}
\begin{equation}
  E_{\rm in}(r,\theta)
  =
  \sum_{n=-N_{\max}}^{N_{\max}}
  b_nJ_n(k_{\text{cyl}}r)\exp(\ii n\theta),
  \quad r < r_{\text{cyl}}.
\label{eqn:E-exact-cyl-vs-rtheta-in}
\end{equation}
Explicitly, the incident plane wave is expanded as
\begin{equation}\label{eqn:incident-cylindrical-expansion}
 E^{\rm inc}_{\alpha}(r,\theta)
 ~=~
 \sum_{n=-\infty}^{\infty}
 c_n(\alpha)J_n(k_{\rm host}r)\exp(\ii n\theta),
 \qquad
 c_n(\alpha)=\ii^n\exp(-\ii n\alpha).
\end{equation}
These coefficients enter the two interface matching equations for each cylindrical harmonic.
The coefficients \(a_n\) and \(b_n\) are determined by the two interface matching equations at \(r=r_{\text{cyl}}\), and the truncation order is selected by a cutoff-convergence check.

\begin{figure}
\centering
\begin{subfigure}[t]{0.48\textwidth}
\centering
\includegraphics[width=\linewidth]{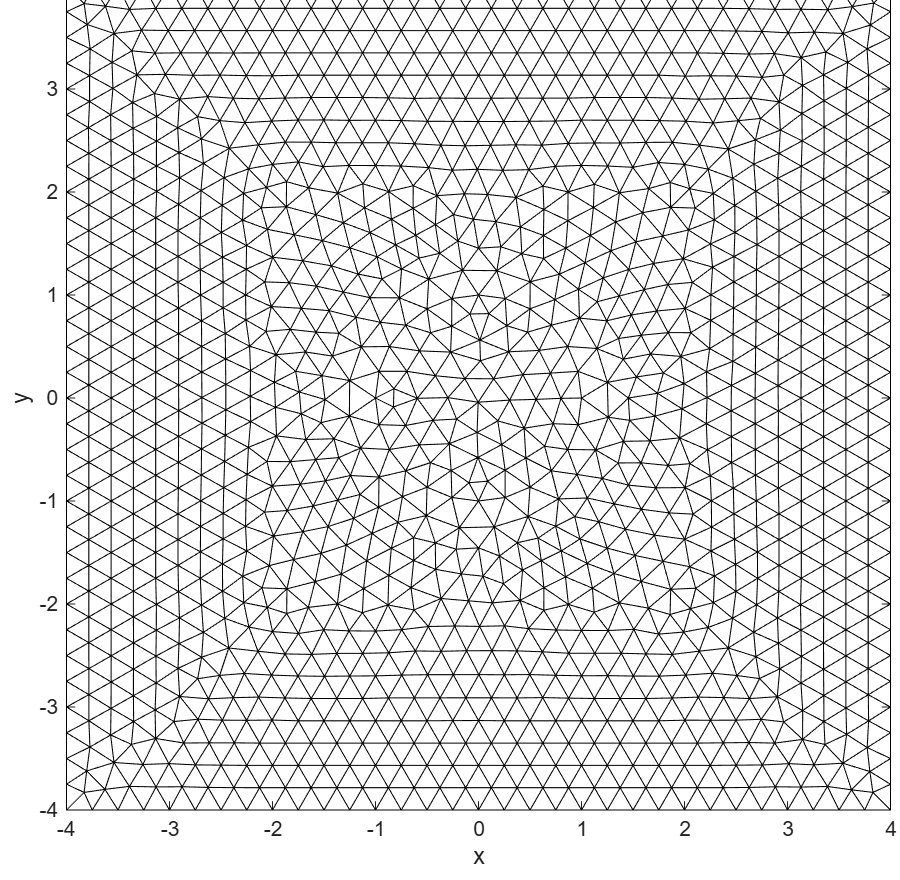}
\caption{Single-cylinder scattering, level 0: 2,302 triangles.}
\end{subfigure}
\hfill
\begin{subfigure}[t]{0.48\textwidth}
\centering
\includegraphics[width=\linewidth]{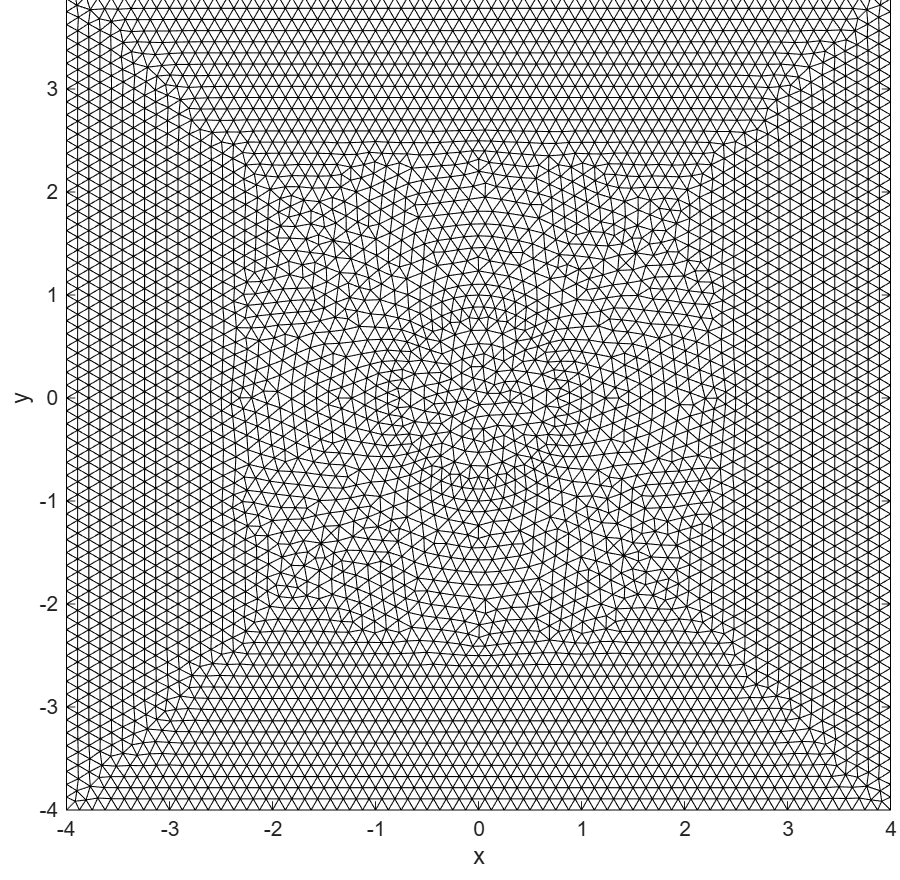}
\caption{Single-cylinder scattering, level 1: 9,138 triangles.}
\end{subfigure}
\caption{Triangular meshes for the single-cylinder scattering example. The computational domain is $[-4,4]^2$. Edge midpoints are not shown.}
\label{fig:single-cylinder-conforming-meshes}
\end{figure}

\paragraph{Meshes}\label{sec:Helmholtz-meshes-assembly-radiation}
For the reported mesh family, \(r_{\rm cyl}=1\), \(\eps_{\rm cyl}=4\), \(\eps_{\rm host}=1\), and \(\Omega=[-4,4]^2\). Scalar standalone FLAME, nodal-triplet GEFLAME, all-triangle $P_2$ FEM, triangular $T$ elements, and triangular TFF are compared on the common level-0, level-1, and level-2 triangular meshes. Level~2 is obtained by uniform red refinement of level~1, with new interface midpoints projected to the circle. All meshes conform to a mesh-aligned piecewise-linear approximation of the circular interface and to the square exterior boundary. Such geometric alignment is not necessary in FLAME but is used here to control the method-to-method comparison. The interface-cut application of FLAME is tested separately on Cartesian grids in \sectref{sec:geflame-cartesian-conforming-comparison}. Table~\ref{table:single-cylinder-mesh-sizes} collects the mesh and algebraic sizes; \figref{fig:single-cylinder-conforming-meshes} shows the two coarser meshes.

Clearly, FEM features a vast range of accuracy-vs-complexity trade-offs: curved or isoparametric elements, higher-order mappings, adaptive $hp$-refinement. These well-known enhancement routes are outside the scope of the present comparison.

\begin{table}
\centering
\caption{Single-cylinder mesh and active algebraic sizes. The scalar-skeleton count applies to scalar FLAME, $P_2$ FEM, the triangular $T$ element, and triangular TFF on all three levels, without the exterior-boundary edge-midpoints.}
\label{table:single-cylinder-mesh-sizes}
\small
\setlength{\tabcolsep}{3pt}
\begin{tabular*}{\textwidth}{@{\extracolsep{\fill}}|c|c|c|c|c|c|c|}
\hline
level & vertices & triangles & \shortstack{exterior\\vertices} & \shortstack{median\\edge} & \shortstack{scalar-skeleton\\DoFs} & \shortstack{GEFLAME\\DoFs} \\
\hline
0 & 1,216 & 2,302 & 128 & 0.250 & 4,605 & 3,392 \\
1 & 4,698 & 9,138 & 256 & 0.125 & 18,277 & 13,582 \\
2 & 18,533 & 36,552 & 512 & 0.0625 & 73,105 & 54,575 \\
\hline
\end{tabular*}
\end{table}

%
\paragraph{Nonlocal two-layer Hankel radiation condition}\label{sec:two-layer-Hankel-radiation}
Radiation conditions are not a primary object in this paper; instead, see \cite{TsukermanTrefftzMachine2014,PaganiniScarabosioHiptmairTsukerman2016} and references there. To focus on approximations inside the domain, all examples below use highly accurate nonlocal two-layer Hankel construction, with outgoing cylindrical harmonics sampled on the two adjacent mesh layers at the square boundary.

More precisely, let $u_b$ and $u_i$ denote field samples on the exterior boundary and on an adjacent interior sample layer, and let \(n_{\max}^{\rm rad}\) denote the Hankel radiation-basis cutoff. The scattered field on these two sets is expanded in outgoing cylindrical harmonics indexed by \(m=-n_{\max}^{\rm rad},\ldots,n_{\max}^{\rm rad}\). Collecting their sampled values gives
\begin{equation}\label{eqn:two-layer-Hankel-samples}
  [F_b]_{j,m}
  \,=\,
  H_m^{(1)}(k_0 r_{b,j}) \exp(\ii m\theta_{b,j}),
  \qquad
  [F_i]_{\ell,m}
  \,=\,
  H_m^{(1)}(k_0 r_{i,\ell}) \exp(\ii m\theta_{i,\ell}),
\end{equation}
where the polar coordinates are measured from the center of the cylinder. The two sample sets need not contain the same number of points. For a scattered-field pair $(v_b, v_i)$, the linear residual map
\begin{equation}\label{eqn:two-layer-radiation-operator}
  \mathcal R(v_b,v_i) ~:=~ v_b-F_bF_i^+v_i
\end{equation}
(where `+' denotes the pseudoinverse) annihilates, up to truncation and least-squares error, the outgoing fields spanned by the selected Hankel modes. Applying this operator to $E-E^{\rm inc}$ gives the total-field radiation condition
\begin{equation}\label{eqn:two-layer-Hankel-condition}
  u_b-F_bF_i^+u_i
  ~=~
  u_b^{\rm inc}-F_bF_i^+u_i^{\rm inc}.
\end{equation}
FLAME systems use one edge-midpoint sample in the inner layer associated with each exterior-boundary vertex, whereas GEFLAME uses the complete adjacent interior-vertex layer. For the reported wavelength study, \(\alpha=\pi/2\) and \(\lambda_0\in\{2.5,5,10\}\), so \(E^{\rm inc}=\exp(\ii k_0y)\) and \(c_n(\alpha)=1\). The comparison field and its scaled derivatives use \(N_{\max}=20\). At the shortest wavelength, the contribution of the first omitted order, \( |n|=21 \), to the interface field or to \(k_0^{-1}\partial_r E\) is well below machine precision. All cases use \(n_{\max}^{\rm rad}=24\), i.e., 49 Hankel modes. This cutoff is distinct from \(N_{\max}\): it controls the least-squares outgoing basis sampled on the square exterior boundary, not the semi-analytical harmonic series. 

For several scatterers (\sectref{sec:four-cylinder-geflame}), the same nonlocal two-layer boundary condition is extended to a multipole basis centered at each scatterer. The relative consistency error of this construction in all numerical experiments of this paper is below $2 \times 10^{-9}$.
%
\paragraph{Discretizations and numerical results}\label{sec:Helmholtz-elements}\label{sec:unstructured-flame-cylinder}\label{sec:Rectangular-cylinder}
%
Scattering from a circular dielectric cylinder is an instructive case  because an exact series expansion exists  \eqref{eqn:E-exact-cyl-vs-rtheta-out}--\eqref{eqn:E-exact-cyl-vs-rtheta-in}. The following methods are compared:
\begin{itemize}
\item \emph{FLAME} uses scalar values on the vertex-and-edge-midpoint graph that conforms geometrically to the circular interface. Each molecule is centered at a graph node and includes the adjacent vertices and edge midpoints. The local basis is truncated, when needed, to one fewer function than the number of sampled DoFs. Homogeneous molecules use equiangular plane waves, while interface molecules use matched circular interface harmonics.
\item \emph{Nodal-triplet GEFLAME} uses $(E,k_0^{-1}\partial_x E,k_0^{-1}\partial_y E)$ at nonboundary vertices and scalar $E$ at exterior-boundary vertices. Each one-ring molecule is centered at a vertex and contains all edge-adjacent vertices; a typical valence-six interior molecule therefore samples $3(6+1)=21$ functionals. The local basis contains 15 plane waves in the homogeneous exterior and 18 interface-matched cylindrical harmonics inside the cylinder, on interface-crossing molecules, and for exterior grid nodes closer to the circular interface than $w_{\rm tr}=3h_0$, where $h_0$ is the coarse-grid edge scale. This physical width is held fixed under $h$-refinement, so the number of graph layers covered by the circular basis increases on finer meshes.
\item \emph{All-triangle $P_2$ FEM} is the conventional quadratic Lagrange-element method on the same mesh.
\item \emph{Triangular $T$ elements} use the local edge-projection construction reviewed in \sectref{sec:T-elements-principles}, with the same six quadratic trace DoFs per triangle as in $P_2$ elements. On each triangle, the field is initially approximated by twelve equiangular plane waves at the local material wavenumber. Their linear combination is least-squares fitted to the six quadratic boundary traces, producing six interior Trefftz basis functions for the element matrices.
\item \emph{Conforming triangular TFF} uses the trace-lift construction of \sectref{sec:FLAME-lift-construction}. Each triangle has a uniform barycentric auxiliary grid with ten subdivisions per edge, denoted TFF-sbd10. The grid contains 66 nodes, of which 30 lie on the boundary and 36 are interior; the interior lift is generated by structured $3\times3$ FLAME molecules clipped to the triangle. 
\end{itemize}
All five systems include the nonlocal Hankel relation \eqref{eqn:two-layer-Hankel-condition}. 

Let $\mathcal V_h$ denote the complete set of mesh vertices and let $\mathcal V_h^\circ$ denote the non-exterior-boundary vertices that carry derivative DoFs in GEFLAME. Define the squared relative field error on all mesh vertices by
\begin{equation}\label{eqn:single-cylinder-field-error}
 e_E^2
 ~:=~
 \frac{\|E_h-E_{\rm ex}\|_{\ell^2(\mathcal V_h)}^2}
   {\|E_{\rm ex}\|_{\ell^2(\mathcal V_h)}^2}.
\end{equation}
Further, define the squared relative gradient error on $\mathcal V_h^\circ$ by
\begin{equation}\label{eqn:single-cylinder-gradient-error}
 e_{\nabla E}^2
 ~=~ e_{k_0^{-1}\nabla E}^2
 ~:=~
 \frac{
 \|\partial_x E_h-\partial_x E_{\rm ex}\|_{\ell^2(\mathcal V_h^\circ)}^2
 +\|\partial_y E_h-\partial_y E_{\rm ex}\|_{\ell^2(\mathcal V_h^\circ)}^2}
 {\|\partial_x E_{\rm ex}\|_{\ell^2(\mathcal V_h^\circ)}^2
 +\|\partial_y E_{\rm ex}\|_{\ell^2(\mathcal V_h^\circ)}^2}.
\end{equation}
The equality on the left is self-evident because the common factor $k_0^{-1}$ cancels out. Because the numerator and denominator use the same sample sets, these relative $\ell^2$ ratios equal relative RMS ratios. Field errors use the complete vertex set $\mathcal V_h$, including exterior-boundary vertices. Gradient errors use $\mathcal V_h^\circ$ because the exterior-boundary GEFLAME unknowns contain only the scalar field.

\begin{figure}
\centering
\makebox[\textwidth][c]{\includegraphics[width=1.05\textwidth]{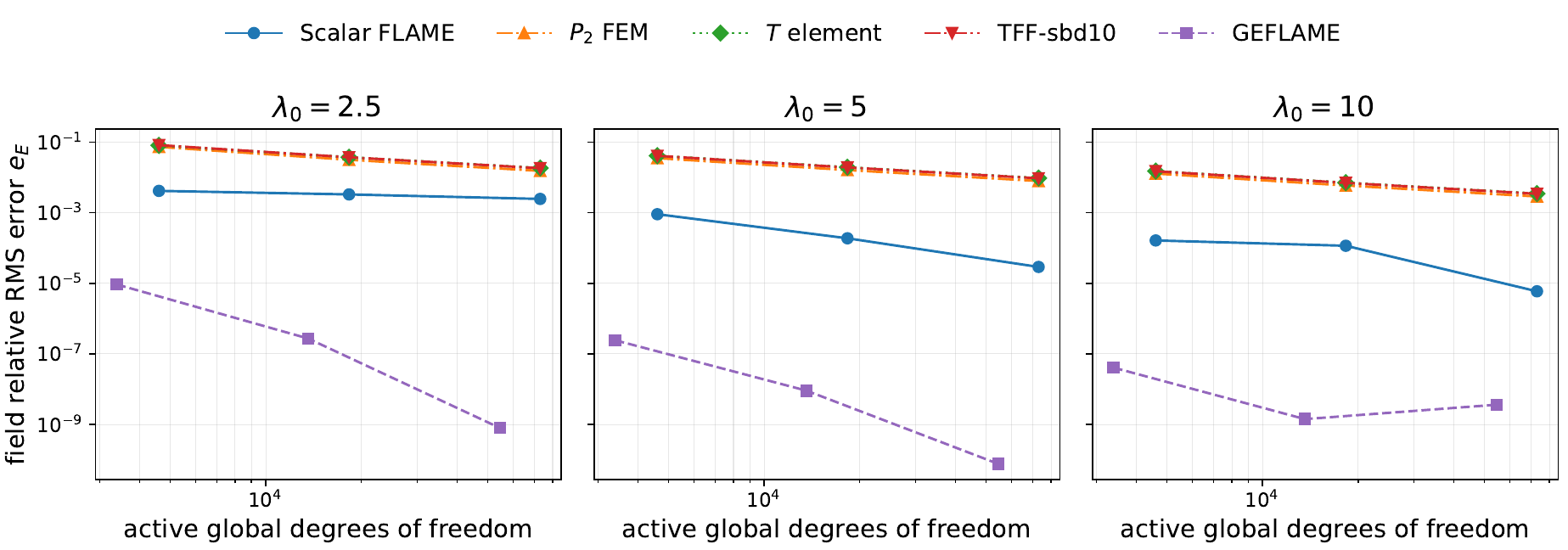}}
\caption{Single-cylinder field error versus the number of active global degrees of freedom. Three points on each line, from left to right, correspond to levels~0, 1, and~2. Field errors are evaluated over the complete mesh-vertex set, including exterior-boundary vertices.}
\label{fig:single-cylinder-error-vs-dofs}
\end{figure}

\begin{figure}
\centering
\makebox[\textwidth][c]{\includegraphics[width=1.05\textwidth]{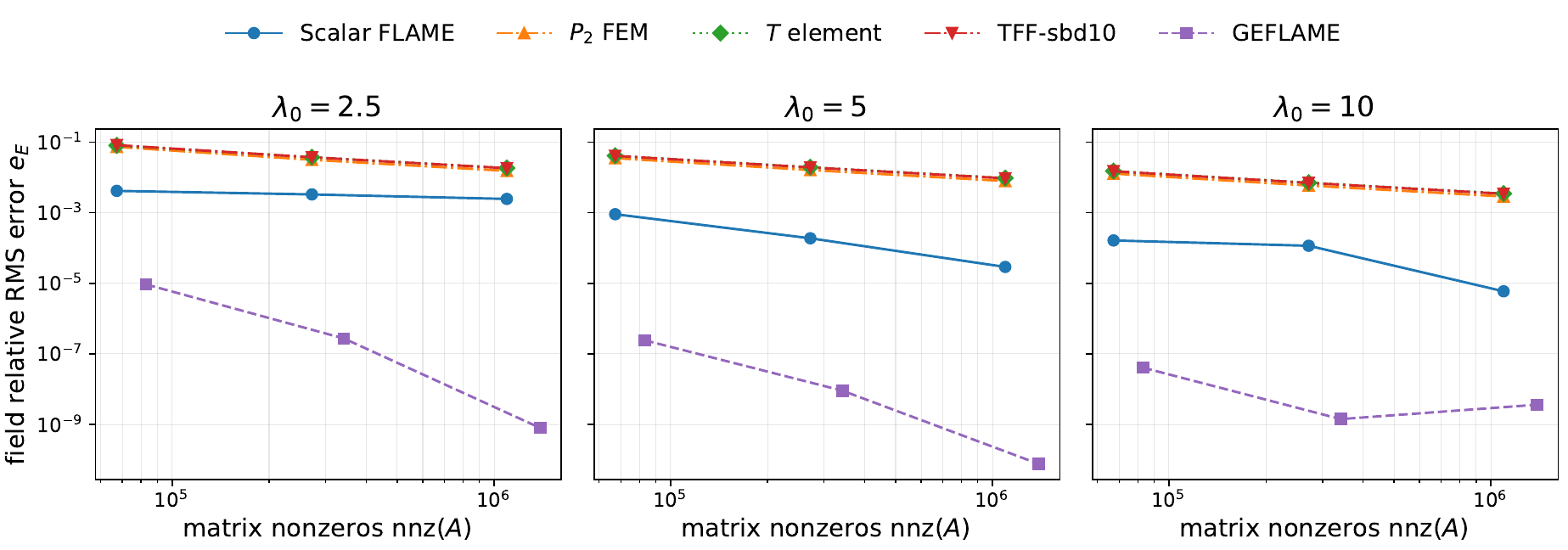}}
\caption{Same as \figref{fig:single-cylinder-error-vs-dofs}, but with errors plotted versus the number of matrix nonzeros.}
\label{fig:single-cylinder-error-vs-nnz}
\end{figure}

Among the four \textit{scalar} discretizations, FLAME has the smallest field error at every wavelength and level. Since the four scalar systems have identical global DoF and nonzero matrix entry counts at each common level,  Figures~\ref{fig:single-cylinder-error-vs-dofs} and~\ref{fig:single-cylinder-error-vs-nnz} provide direct cost-matched accuracy comparisons for the implementations tested. FLAME lowers the field error by factors of about 20--90 on level~0, 10--100 on level~1, and 6--600 on level~2, for the three wavelengths and three comparison methods -- $P_2$,  $T$ and TFF.

The nodal-triplet GEFLAME calculation gives substantially smaller field errors and also supplies native gradients with comparable accuracy. Relative to FLAME, already the most accurate scalar method in this comparison, the GEFLAME field errors are lower \textit{from two to six orders of magnitude} over the nine wavelength/level pairs shown in \figref{fig:single-cylinder-error-vs-nnz}. Notably, at every wavelength and mesh level in \tableref{table:single-cylinder-geflame-results}, the gradient and field errors remain within a factor of $\sim$1--3 of one another; in the present comparisons, this is a notable feature of GEFLAME. On all three levels, GEFLAME uses fewer global unknowns than the scalar-skeleton methods because the global DoF sets are not the same. Each scalar-skeleton method places one scalar DoF at every mesh vertex and every edge midpoint, for a total of $V+E$ unknowns on a triangular mesh. GEFLAME instead places a triplet only at non-exterior-boundary vertices and a single field DoF at exterior-boundary vertices, giving approximately $3V$ unknowns. Since $E$ is close to $3V$ on these meshes, $V+E$ is approximately $4V$ and exceeds the GEFLAME count despite the derivative enrichment. The GEFLAME matrices nevertheless contain about 25\% more nonzeros. The field-error/cost comparisons are shown against global DoFs in \figref{fig:single-cylinder-error-vs-dofs} and against matrix nonzeros in \figref{fig:single-cylinder-error-vs-nnz}; the numerical field and gradient errors are collected in \tableref{table:single-cylinder-geflame-results}.

\begin{table}
\centering
\small
\setlength{\tabcolsep}{3pt}
\caption{Nodal-triplet GEFLAME field and gradient relative RMS errors. Field errors use the complete mesh-vertex set; gradient errors use the non-exterior-boundary triplet vertices.}
\label{table:single-cylinder-geflame-results}
\begin{tabular*}{\textwidth}{@{\extracolsep{\fill}}|r|r|r|r|r|r|r|}
\hline
\multicolumn{1}{|c|}{} & \multicolumn{3}{c|}{$e_E$} & \multicolumn{3}{c|}{$e_{\nabla E}$} \\
\hline
$\lambda_0$ & level 0 & level 1 & level 2 & level 0 & level 1 & level 2 \\
\hline
2.5 & $9.26\times10^{-6}$ & $2.73\times10^{-7}$ & $7.97\times10^{-10}$ & $1.09\times10^{-5}$ & $3.26\times10^{-7}$ & $1.08\times10^{-9}$ \\
5 & $2.43\times10^{-7}$ & $9.08\times10^{-9}$ & $7.64\times10^{-11}$ & $4.41\times10^{-7}$ & $1.31\times10^{-8}$ & $1.53\times10^{-10}$ \\
10 & $4.07\times10^{-8}$ & $1.42\times10^{-9}$ & $3.63\times10^{-9}$ & $8.85\times10^{-8}$ & $2.33\times10^{-9}$ & $1.15\times10^{-8}$ \\
\hline
\end{tabular*}
\end{table}

At $\lambda_0=10$, the level-2 GEFLAME errors rise modestly after reaching the $10^{-9}$ scale on level~1, but remain at or below $10^{-8}$. This indicates that the calculation has reached an error floor rather than continuing in a clean asymptotic regime.
%
\FloatBarrier
\subsection{Four-cylinder scattering example}\label{sec:four-cylinder-geflame}
%
\paragraph{Semi-analytical multipole-multicenter solution}
The governing $s$-mode equation, interface conditions, and scattering conventions are those of \sectref{sec:Helmholtz-problem-formulation}.
For several dielectric circular cylinders, the quasi-exact solution of $s$-mode scattering can be written as the standard multipole-multicenter expansion for parallel circular cylinders \cite{Twersky1952ParallelCylinders,TsueiBarber1988}. For $J$ circular cylinders and the $s$-mode (one-component electric field), let $n_{\max}^{\rm mp}$ denote the multipole truncation. Let $\Omega_j$ be the interior of cylinder $j$ and let $\Omega_{\rm host}$ be the surrounding free-space region. The total field in the host is
\begin{equation}\label{eqn:four-cylinder-mmc-host}
E^{\rm host}(\bfr)
= E^{\rm inc}(\bfr)
 + \sum_{j=1}^{J}\sum_{n=-n_{\max}^{\rm mp}}^{n_{\max}^{\rm mp}}
   a_{j,n} H_n^{(1)}(k_0 r_j) \exp(\ii n\theta_j),
\qquad \bfr\in\Omega_{\rm host}.
\end{equation}
The field inside cylinder $j$ is
\begin{equation}\label{eqn:four-cylinder-mmc-cylinder}
E_j^{\rm cyl}(\bfr)
= \sum_{n=-n_{\max}^{\rm mp}}^{n_{\max}^{\rm mp}}
   b_{j,n} J_n(k_j r_j)\exp(\ii n\theta_j),
\qquad \bfr\in\Omega_j,
\qquad k_j=k_0\sqrt{\eps_j}.
\end{equation}
Here $(r_j,\theta_j)$ are polar coordinates centered at cylinder $j$. The coefficients $a_{j,n}$ and $b_{j,n}$ are determined by enforcing $[E]=0$ and $[\partial_n E]=0$ on each circular interface. Contributions centered at other cylinders, as well as the incident field, are re-expanded about cylinder $j$ by Graf's addition theorem \cite{Twersky1952ParallelCylinders,TsueiBarber1988}. The tested configuration has $J=4$ cylinders, with the geometry and material data listed in \tableref{table:four-cylinder-geometry}, where $(c_x, c_y)$ are coordinates of the cylinder centers.

\begin{table}
\centering
\caption{Four-cylinder geometry and material data}
\label{table:four-cylinder-geometry}
\begin{tabular}{|c|c|c|c|c|}
\hline
cylinder & $c_x$ & $c_y$ & radius & $\eps$ \\
\hline
C1 & $0.08$ & $-2.34$ & $0.8025$ & $1.787$ \\
\hline
C2 & $-0.03$ & $2.87$ & $0.9900$ & $4.186$ \\
\hline
C3 & $-1.61$ & $0.12$ & $0.5250$ & $3.356$ \\
\hline
C4 & $1.83$ & $-0.35$ & $0.6750$ & $3.840$ \\
\hline
\end{tabular}
\end{table}

For the reported mesh ladder and scattering calculations, the free-space wavelength is normalized to \(\lambda_0=1\), so \(k_0=2\pi\), the computational box is \([-4.5,4.5]\times[-5.5,5.5]\), and the incident field is \(E^{\rm inc}(x,y)=\exp(\ii k_0y)\). The two-layer Hankel radiation condition of \sectref{sec:two-layer-Hankel-radiation} is imposed on the exterior boundary.

\begin{figure}
\centering
\begin{subfigure}{0.48\linewidth}
\centering
\includegraphics[width=\linewidth]{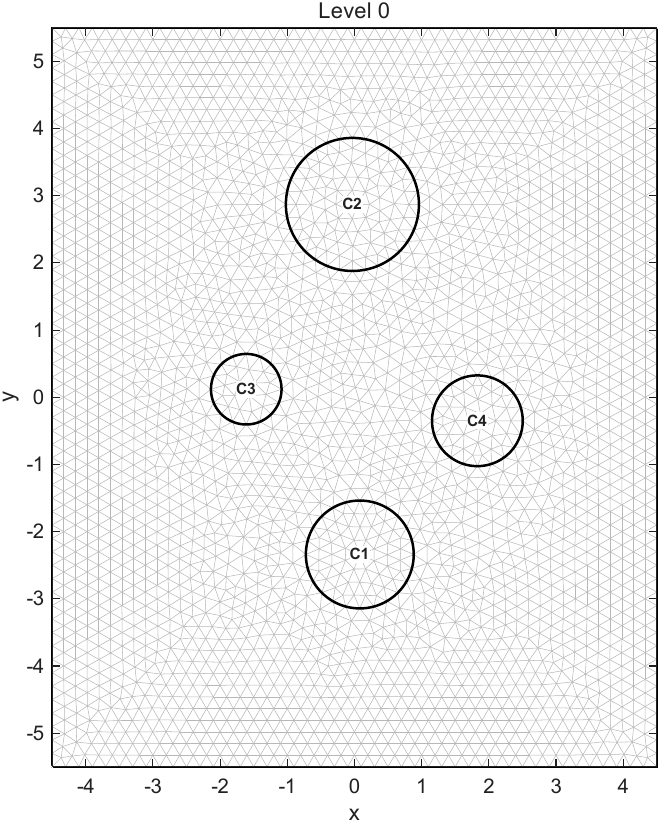}
\caption{Four-cylinder scattering, level 0.}
\end{subfigure}\hfill
\begin{subfigure}{0.48\linewidth}
\centering
\includegraphics[width=\linewidth]{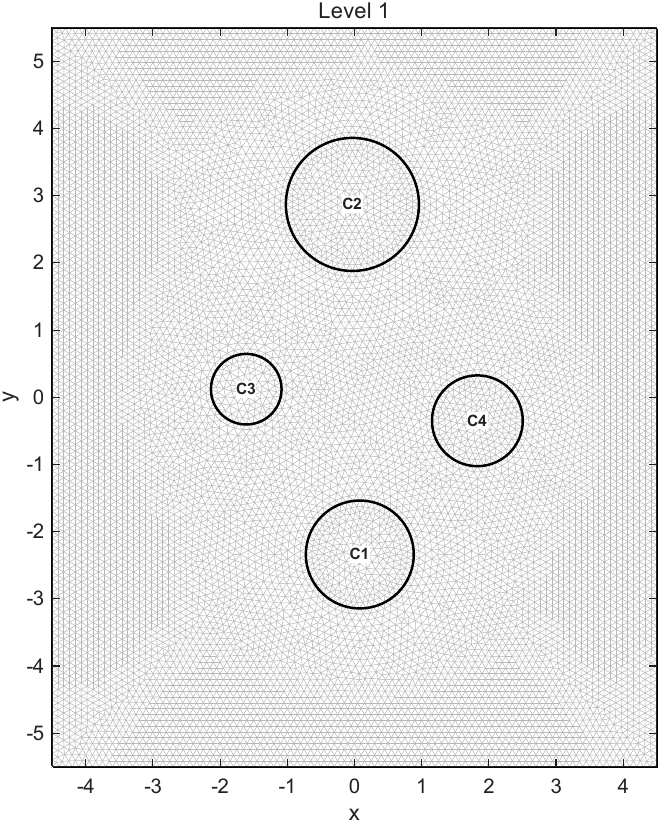}
\caption{Four-cylinder scattering, level 1.}
\end{subfigure}
\caption{Four-cylinder scattering example: triangular meshes aligned with piecewise-linear approximations of the circular interfaces.}
\label{fig:four-cylinder-meshes}
\end{figure}

\tableref{table:four-cylinder-mesh-ladder} includes data for a mesh ladder aligned with piecewise-linear approximations of the four circular interfaces; curved elements are not used. \figref{fig:four-cylinder-meshes} shows the first two levels. As in the single-cylinder test, the GEFLAME bases encode the exact circular interfaces, whereas the tested $P_2$ implementation uses the mesh-aligned piecewise-linear material boundaries.

\begin{table}
\centering
\caption{Data for three levels of the four-cylinder mesh ladder; ppw = points per wavelength.}
\label{table:four-cylinder-mesh-ladder}
\begin{tabular}{|r|r|r|r|r|r|}
\hline
level & ppw & $h_{\max}$ & vertices & triangles & median edge \\
\hline
0 & 5 & 0.2000 & 2,860 & 5,518 & 0.2000 \\
\hline
1 & 10 & 0.1000 & 11,000 & 21,598 & 0.1000 \\
\hline
2 & 20 & 0.0500 & 44,908 & 89,014 & 0.0500 \\
\hline
\end{tabular}
\end{table}

\paragraph{GEFLAME}
The four-cylinder calculation uses the same nodal-triplet GEFLAME construction as in the single-cylinder case (\sectref{sec:Helmholtz-elements}): one-ring molecules, 15 plane waves in homogeneous regions, and 18 circular interface-matched functions inside or near each cylinder. The cylinders are quasi-isolated at the scale $w_{\rm tr}=3h_0=0.60$: every interface-near molecule is associated unambiguously with one cylinder, while the physical transition width is held fixed under refinement.

Because a one-ring molecule samples more functionals than the 15- or 18-function local basis, its nullspace is generally multidimensional, as it is in the single-cylinder nodal-triplet calculation. The three schemes associated with the field and two derivative DoFs at the central vertex are selected by heuristic stability optimization \eqref{eqn:flame-multidimensional-scheme-selection}.

\begin{figure}
\centering
\includegraphics[width=0.88\textwidth]{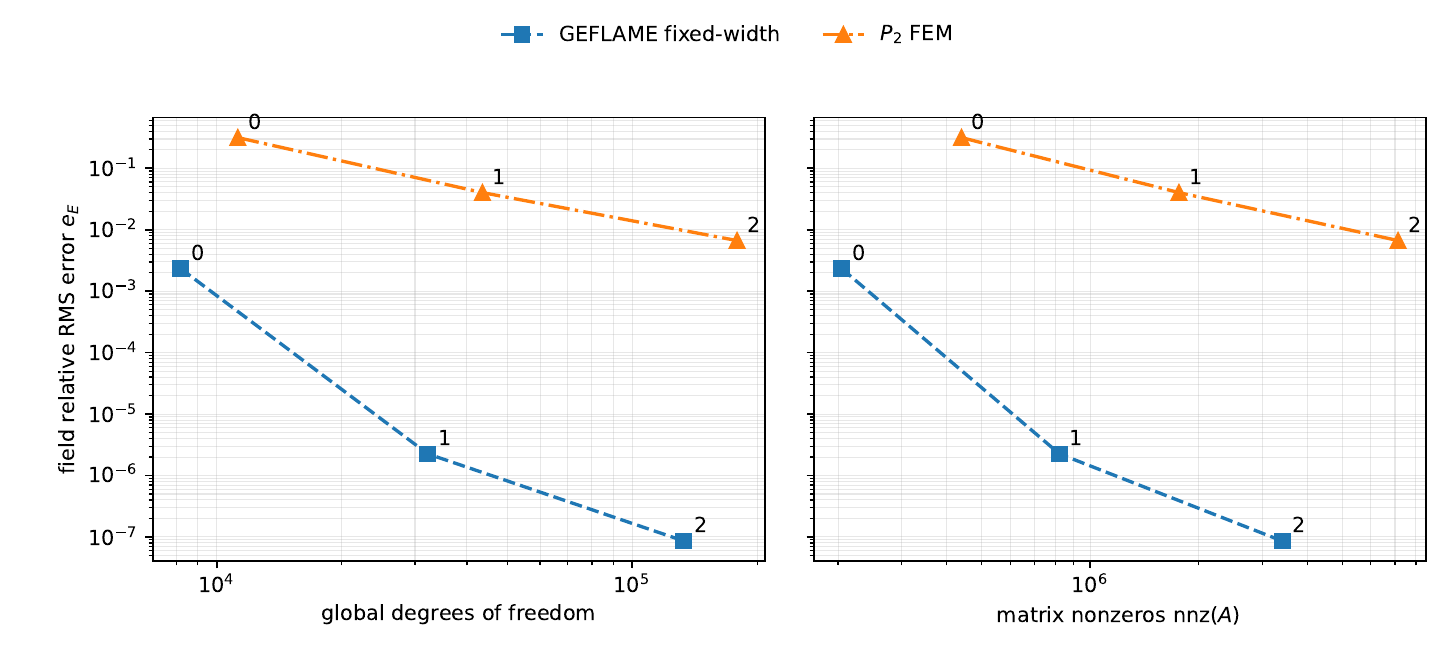}
\caption{Four-cylinder field relative RMS error on the common 701-point vertical midline versus global degrees of freedom and matrix nonzeros. Numerals next to the markers denote mesh levels~0--2. The system-size data are reported in \tableref{table:four-cylinder-results}.}
\label{fig:four-cylinder-error-vs-cost}
\end{figure}

\paragraph{Accuracy and cost}
The multipole comparison field uses \(n_{\max}^{\rm mp}=32\); increasing the cutoff does not materially change the sampled fields or derivatives. For a method-independent comparison, both GEFLAME and $P_2$ field and gradient errors are evaluated on the same 701-point vertical midline, $x=0$ and $-5.25 \le y \le 5.25$. GEFLAME values are reconstructed from the native triplet solution and $P_2$ values from the quadratic finite-element interpolation.
These data are plotted in \figref{fig:four-cylinder-error-vs-cost}. In addition,  Table~\ref{table:four-cylinder-results} records the system sizes. Both methods are compared with the $n_{\max}^{\rm mp}=32$ multipole-multicenter solution.

\begin{table}
\centering
\caption{Four-cylinder system sizes and field and gradient relative RMS errors on the common 701-point vertical midline. The same data are shown against both degrees of freedom and matrix nonzeros in Figures~\ref{fig:four-cylinder-error-vs-cost} and~\ref{fig:four-cylinder-gradient-error-vs-cost}. ppw = points per wavelength.}
\label{table:four-cylinder-results}
\small
\setlength{\tabcolsep}{3pt}
\begin{tabular*}{\textwidth}{@{\extracolsep{\fill}}|l|c|c|c|c|c|c|}
\hline
method & level & ppw & DoFs & $\operatorname{nnz}(A)$ & $e_E$ & $e_{\nabla E}$ \\
\hline
GEFLAME & 0 & 5 & 8,180 & 204,586 & $2.33\times10^{-3}$ & $2.83\times10^{-3}$ \\
GEFLAME & 1 & 10 & 32,200 & 821,806 & $2.25\times10^{-6}$ & $2.36\times10^{-6}$ \\
GEFLAME & 2 & 20 & 133,124 & 3,406,810 & $8.60\times10^{-8}$ & $8.71\times10^{-8}$ \\
$P_2$ FEM & 0 & 5 & 11,237 & 439,727 & $3.16\times10^{-1}$ & $3.62\times10^{-1}$ \\
$P_2$ FEM & 1 & 10 & 43,597 & 1,762,367 & $4.03\times10^{-2}$ & $5.11\times10^{-2}$ \\
$P_2$ FEM & 2 & 20 & 178,829 & 7,138,535 & $6.67\times10^{-3}$ & $1.00\times10^{-2}$ \\
\hline
\end{tabular*}
\end{table}

\begin{figure}
\centering
\includegraphics[width=0.88\textwidth]{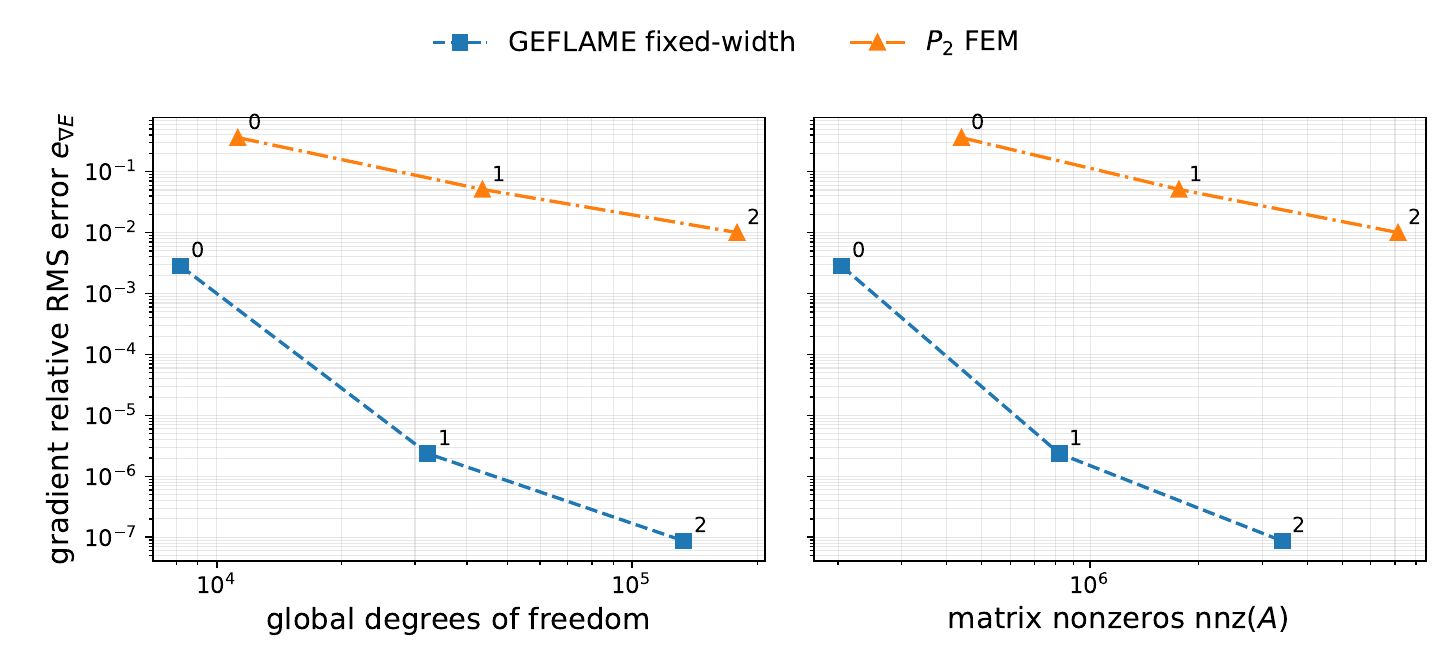}
\caption{Four-cylinder gradient relative RMS error on the common 701-point vertical midline versus global degrees of freedom and matrix nonzeros. Numerals next to the markers denote mesh levels~0--2. The system-size data are reported in \tableref{table:four-cylinder-results}.}
\label{fig:four-cylinder-gradient-error-vs-cost}
\end{figure}

The four-cylinder cost--accuracy comparisons in Figures~\ref{fig:four-cylinder-error-vs-cost} and~\ref{fig:four-cylinder-gradient-error-vs-cost} show a clear advantage for GEFLAME in this test problem. Its midline field error decreases from $2.33\times10^{-3}$ to $2.25\times10^{-6}$ and $8.60\times10^{-8}$ under $h$-refinement, while its gradient error decreases from $2.83\times10^{-3}$ to $2.36\times10^{-6}$ and $8.71\times10^{-8}$. At the three levels, the GEFLAME field errors are lower than the $P_2$ errors by factors of approximately $1.4\times10^2$, $1.8\times10^4$, and $7.8\times10^4$; the corresponding gradient-error factors are approximately $1.3\times10^2$, $2.2\times10^4$, and $1.1\times10^5$. In short, both errors are about two to five orders of magnitude lower in GEFLAME.

The established FEM enhancement routes noted in \sectref{sec:Helmholtz-meshes-assembly-radiation} remain available; \tableref{table:four-cylinder-results} and Figures~\ref{fig:four-cylinder-error-vs-cost} and~\ref{fig:four-cylinder-gradient-error-vs-cost} compare only the two implementations described here.
\FloatBarrier
\subsection{Nodal-triplet vs node--edge GEFLAME on the four-cylinder test}\label{sec:geflame-dof-layout-comparison}
%
FLAME depends on the choice of local Trefftz approximation space and degrees of freedom (DoFs). In the nodal-triplet calculation each interior mesh vertex carries
\begin{equation}
    \mathcal{D}_{\rm nodal}(u;v)
    \,=\,
    \bigl(u(v), ~ k_0^{-1} \partial_x u(v), 
    ~ k_0^{-1}\partial_y u(v)\bigr).
\label{eqn:nodal-triplet-dofs}
\end{equation}
with the physically natural $k_0^{-1}$ normalization. 

An alternative node--edge set of DoFs retains the scalar values at vertices but replaces nodal gradients by edge-normal derivative DoFs at edge midpoints,
\begin{equation}\label{eqn:node-edge-dofs}
    \mathcal{D}_{\rm edge}^{s}(E; e)
    ~=~
    |e| \, \nabla E(\bfr_e) \cdot \bfn_e,
\end{equation}
where $\bfr_e$ is the edge midpoint and $\bfn_e$ is the edge normal in one of the two possible directions. The node--edge formulation is closer in spirit to flux and circulation DoFs \cite{Nedelec1980,Nedelec1986,Bossavit1998}, although GEFLAME remains a nonconforming finite-difference construction rather than an $H(\operatorname{curl})$-conforming finite element method.

For a fair comparison, the same local approximation spaces and the same fixed transition width $w_{\rm tr}=0.60$ are used in both variants: 15 plane waves in homogeneous exterior regions and 18 circular interface modes inside a cylinder, on interface-crossing molecules, and throughout its transition neighborhood. A nodal molecule uses the one-ring vertex samples described above; the corresponding node--edge molecule uses scalar samples at those vertices and normal-derivative samples at the midpoints of the incident edges. Both variants use the generalized-eigenvalue selection \eqref{eqn:flame-multidimensional-scheme-selection}.

At ppw~10, using the common all-vertex field norm, the nodal-triplet relative RMS error is $2.85\times10^{-6}$, whereas the node--edge error is $2.72\times10^{-5}$. Thus the nodal-triplet calculation is about 9.5 times more accurate while using approximately 25\% fewer DoFs and 52\% fewer nonzeros. For this test problem, co-locating the first derivatives with the field at vertices is therefore both more accurate and more economical than distributing normal derivatives to edges.
%
\subsection{Geometrically conforming and Cartesian nodal-triplet GEFLAME}\label{sec:geflame-cartesian-conforming-comparison}
%
The same $s$-mode four-cylinder problem is also solved on regular Cartesian grids with grid sizes equal to the average level-0 and level-1 triangular-mesh edge lengths. Both discretizations are nodal-triplet GEFLAME: the unknowns are $(E,k_0^{-1}\partial_x E,k_0^{-1}\partial_y E)$ away from the exterior boundary. On the Cartesian grids, FLAME is constructed from fixed $3\times3$ node molecules; away from the boundary, each molecule contains nine nodes and therefore 27 sampled DoFs. On triangular meshes, FLAME uses one-ring mesh molecules, which gives on average 21 sampled values for a typical node with 6 neighbors. Trefftz bases in both discretizations consist of 15 homogeneous plane waves away from the cylinders and 18 exact circular interface modes inside the cylinders, on interface-crossing molecules, and throughout the same fixed physical transition width $w_{\rm tr}=0.60$. Both use the same two-layer nonlocal Hankel exterior radiation condition. 

\tableref{table:cartesian-vs-conforming-global} reports the results. The Cartesian-grid systems have about 8--10\% fewer unknowns but over 10\% more nonzeros. On each geometry's native nodal set, the Cartesian field errors are larger by factors of roughly 2 and 11 at the coarse and fine scales, respectively.

\begin{table}
\centering
\small
\caption{Four-cylinder $s$-mode nodal-triplet GEFLAME on geometrically conforming triangular meshes and regular Cartesian grids. Both formulations use $(E,k_0^{-1}\partial_x E,k_0^{-1}\partial_y E)$. The Cartesian spacings $h=0.2$ and $h=0.1$ correspond to levels~0 and~1. The scalar relative RMS error $e_E$ is measured against the same $n_{\max}^{\rm mp}=32$ multiple-cylinder comparison solution and is evaluated on each geometry's native nodal set.}
\label{table:cartesian-vs-conforming-global}
\begin{tabular}{|c|c|c|c|c|}
\hline
scale & sampling geometry & DoFs & nnz$(A)$ & $e_E$ \\
\hline
level 0 / $h\simeq0.2$ & conforming triangular & 8,180 & 204,586 & $1.93\times10^{-2}$ \\
\hline
level 0 / $h=0.2$ & Cartesian $3\times3$ & 7,328 & 227,552 & $3.49\times10^{-2}$ \\
\hline
level 1 / $h\simeq0.1$ & conforming triangular & 32,200 & 821,806 & $2.85\times10^{-6}$ \\
\hline
level 1 / $h=0.1$ & Cartesian $3\times3$ & 29,503 & 935,877 & $3.26\times10^{-5}$ \\
\hline
\end{tabular}
\end{table}

A separate local diagnostic helps explain the counter-intuitive relatively poor performance of the schemes on regular grids. Statistically representative near-interface grid molecules were sampled with the same one-cylinder interface-matched spaces for eight incident test waves with directions separated by $45^\circ$. For each exact sampled DoF vector $d(u)$ and each selected central-DoF scheme $s_i$, the normalized consistency error was
\begin{equation}\label{eqn:cartesian-conforming-local-consistency}
    \eta_i
    ~=~
    \frac{|s_i^\top d(u)|}{\|s_i\|_2\,\|d(u)\|_2}.
\end{equation}
The median and 90th-percentile consistency errors at level~0 are smaller for geometrically conforming sampling by factors between 5 and 6; at level~1 the corresponding factors are over 30. 

Regular Cartesian FLAME molecules are still preferable in homogeneous regions, where the order of the scheme may become higher by serendipity \cite{Tsukerman2005,Tsukerman2006,TsukermanCajko2008}.
\FloatBarrier
\subsection{Reentrant-corner singularity: localized enrichment}\label{sec:Reentrant-corner-test}
\begin{figure}
\centering
\includegraphics[width=0.5\linewidth]{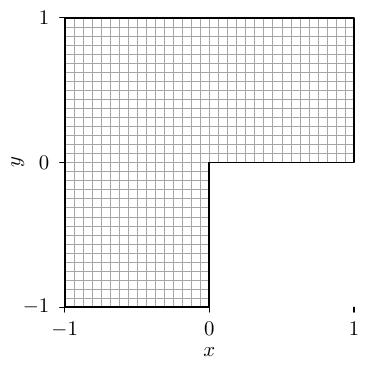}
\caption{The L-shaped computational domain $\Omega_L$ and the coarsest Cartesian grid, $N=32$.}
\label{fig:reentrant-corner-domain-mesh}
\end{figure}

%
This example is inspired by the L-shaped-domain Helmholtz test in \cite{LieuGabardBeriot2016} and is used here to examine singular-function enrichment; related corner-adapted FLAME for scattering by perfectly conducting objects are described in \cite{AlKhateeb13}. 
The L-shaped computational domain is (\figref{fig:reentrant-corner-domain-mesh})

\begin{equation}
  \Omega_L \,=\, [-1,1]^2 \setminus \bigl((0,1] \times [-1,0) \bigr),
\end{equation}
so the origin is a reentrant corner with aperture $3\pi/2$. The fixed-frequency scalar Helmholtz equation, $\nabla^2E+k^2E=0$, has the manufactured exact field
\begin{equation}\label{eqn:reentrant-corner-exact}
  E_{\rm ex}(r,\theta)
  \,=\,
  J_{2/3}(kr)\sin\!\left(\frac{2\theta}{3}\right),
  \qquad 0<\theta<\frac{3\pi}{2}.
\end{equation}
The field is bounded, but its gradient is singular at the corner. Homogeneous Dirichlet conditions are imposed on the two reentrant rays. On the outer square boundary, local outgoing-wave FLAME equations are used, with their right-hand sides evaluated from \eqref{eqn:reentrant-corner-exact}. Thus the exact field satisfies the boundary equations by construction, and the experiment isolates the accuracy of the interior approximation.

A uniform Cartesian grid with spacing $h=2/N$ is clipped to $\Omega_L$, and each interior equation is associated with a $3\times3$ molecule. Outside the enriched corner patch, the local Trefftz basis contains eight propagating plane waves with directions $\theta_j=j\pi/4$, $j=0,\ldots,7$. Inside the patch, the basis contains seven equally spaced propagating plane waves plus the fractional-Bessel function in \eqref{eqn:reentrant-corner-exact}, anchored at the reentrant corner. Both local bases therefore contain eight functions sampled at nine nodes and generically produce a one-dimensional nullspace of the transpose sampling matrix. 

The corner patch is defined in a natural way: a molecule centered at $(i,j)$ is enriched when
\begin{equation}
  \max \bigl(|i-i_c|, \, |j-j_c|\bigr) \,\le\, m,
\end{equation}
where $(i_c,j_c)$ is the corner index and $m$ is the adjustable patch radius in grid layers.

In the example, $kL=50$ with $L=|\Omega_L|^{1/2}=\sqrt{3}$; $N=96$, hence $h=1/48$, giving $7\,105$ nodal unknowns. Parameter $m$ varies from 0 through 47; $m=47$ is the first value for which all interior molecules are enriched.

\begin{figure}
\centering
\includegraphics[width=0.82\linewidth]{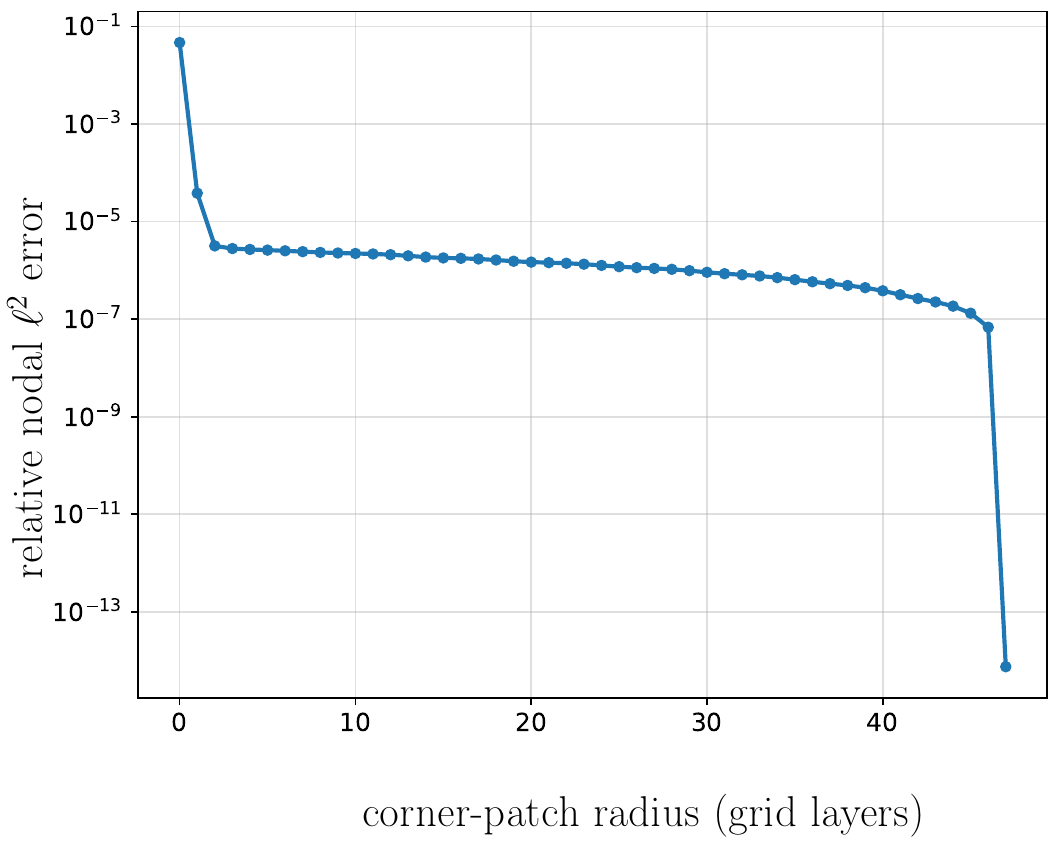}
\caption{Reentrant-corner localized-enrichment example: relative nodal $\ell^2$ error versus corner-patch radius in grid layers. The first two enriched layers rectify the dominant corner defect. The final drop at $m=47$ occurs when the manufactured field is included in every local basis.}
\label{fig:reentrant-corner-convergence}
\end{figure}

For the vectors of nodal field samples, the reported relative error is
\begin{equation}
 e_2 ~=~ \frac{\|E_h-E_{\rm ex}\|_2}
              {\|E_{\rm ex}\|_2}.
\end{equation}
\figref{fig:reentrant-corner-convergence} displays this error as a function of the number of layers in the corner patch. The first enriched layer lowers $e_2$ by more than three orders of magnitude, and the second layer gives a further factor of about $12$. From $m=2$ through $m=46$, the error then decreases relatively slowly (but note the logarithmic scale). The final error drop occurs when the exact fractional-Bessel field gets included in every interior local basis.

A trend-level comparison with \cite{LieuGabardBeriot2016} is meaningful because the same L-shaped corner field and the same $kL$ normalization are used. That study shows that the corner singularity destroys the high-order convergence otherwise expected under uniform $p$-refinement, while local $h$-refinement improves both the polynomial and wave-based methods. The present calculation reaches the analogous conclusion by a different mechanism: on a fixed mesh, two layers of singular-function enrichment reduce $e_2$ from $4.659\times10^{-2}$ to $2.794\times10^{-6}$, a factor of approximately $1.7\times10^4$. A cross-paper error comparison would nevertheless be unreliable. The study in \cite{LieuGabardBeriot2016} uses unstructured triangular $p$-FEM and wave-based DG discretizations, a quadrature-based continuous $L^2(\Omega_L)$ norm, different Robin-type outer-boundary conditions, and a degrees-of-freedom-per-wavelength abscissa $D_\lambda$. The present calculation instead uses a Cartesian grid, the discrete nodal norm $e_2$, and manufactured outgoing-wave FLAME boundary equations.
%
\FloatBarrier
\section{Computation of Bloch bands}\label{sec:Bloch-tests}
%
\subsection{Motivation}\label{sec:Bloch-bands-motivation}
Bloch eigenvalue problems are central in the analysis of periodic media, including photonic crystals, phononic and acoustic crystals, periodic composites, and electromagnetic metamaterials. The resulting band diagram identifies pass bands, stop bands, mode crossings, and is often the first computational object used to understand wave propagation in a periodic material.

Such calculations are demanding because one usually solves many unit-cell problems along a path in the Brillouin zone, and because the modes of interest may be propagating or evanescent. The Bloch variety -- the graph \( (\bfq, \omega) \) -- is in practice sampled as $\omega(\bfq)$ or, alternatively, as $\bfq(\omega)$. The latter formulation is especially useful for evanescent, complex, or lossy modes, and is also natural for Trefftz bases, which are built at a given frequency. 
%
\subsection{Formulation}\label{sec:Bloch-formulation}
%
Consider a rectangular periodic cell $\Omega = [-a_x/2,a_x/2] \times [-a_y/2,a_y/2]$ with periodic relative permittivity $\eps(x,y)$. The one-component electric field satisfies
\begin{equation}\label{eqn:Bloch-Helmholtz}
   \nabla^2 E \,+\, k_0^2 \eps(x,y) E \,=\, 0,
   \qquad k_0=\frac{\omega}{c},
\end{equation}
with $[E] = 0$, $[\partial_n E] = 0$ across material interfaces. For a square cell of side $a$, the standard Bloch conditions are
\begin{equation}\label{eqn:Bloch-bc}
   E(x+a,y) \,=\, z_xE(x,y),
   \quad E(x, y+a) \,=\, z_yE(x,y),
   \quad z_j = \exp(\ii q_ja).
\end{equation}
The $q(\omega)$ problem can be formulated either in terms of the full field (FF), with Bloch multipliers imposed between opposite cell boundaries, or in terms of the lattice-periodic factor (PF), as in established formulations such as \cite{LacknerMengMonk2019,HuangLiLinLinTian2020,KlindworthSchmidtFliss2014}. The two descriptions are equivalent at the continuous level. After discretization, however, their algebraic structures can differ qualitatively \cite{TsukermanFullFieldPeriodic2026}. In the FF formulation, the Bloch eigenvalue enters through $z_j=\exp(\ii q_ja)$ only in the opposite-boundary coupling, so reciprocal-lattice shifts leave the discrete problem unchanged. Standard PF discretizations instead place components of $\bfq$ in the differential operator; this can break reciprocal-lattice (Brillouin-zone-shift) covariance and may produce inaccurate or phantom modes in $q(\omega)$ calculations \cite{TsukermanFullFieldPeriodic2026}. This does not rule out suitably designed PF formulations, nor does it call into question the conventional use of PF at prescribed $q$ in $\omega(q)$ calculations; it motivates the use of the FF formulation for the present $q(\omega)$ problem.
%
\subsubsection{Sesquilinear and bilinear Bloch forms}\label{sec:Sesquilinear-vs-bilinear}
%
The paired integration-by-parts contributions from the opposite cell edges, if retained in the variational formulation, would introduce additional $z$-dependent terms and complicate the eigenvalue problem. Here, the basis and test functions are defined in such a way that these paired contributions cancel.

The standard Hilbert-space form is sesquilinear,
\begin{equation}\label{eqn:Bloch-sesquilinear}
 a(E,W)=\int_\Omega \nabla E \cdot \nabla W^* \, \dd \bfr
     - k_0^2 \int_\Omega \eps \, E W^* \dd \bfr,
\end{equation}
whereas the bilinear alternative is
\begin{equation}
 b(E,W)=\int_\Omega \nabla E\cdot\nabla W\,\dd\bfr
 -k_0^2\int_\Omega \eps E W\,\dd\bfr.
\label{eqn:Bloch-bilinear}
\end{equation}
Let a basis function satisfy $E_+=z_jE_-$ on a pair of opposite faces. After accounting for the opposing outward normals, the paired sesquilinear boundary residual is proportional to $z_jW_+^*-W_-^*$. Cancellation therefore requires
\begin{equation}\label{eqn:Bloch-adjoint-test-phase}
    W_+ ~=~  {z_j^*}^{-1} W_- .
\end{equation}
\paragraph{Sesquilinear form on the unit circle}
For a lossless reciprocal problem, $|z_j|=1$, and ${z_j^*}^{-1}=z_j$. The basis and test functions then belong to the same Bloch space, giving the standard Galerkin setting and the expected Hermitian structure.

\paragraph{Sesquilinear form off the unit circle}
If $|z_j|\ne1$, a same-space sesquilinear formulation fails to cancel the paired periodic-face terms and introduces the nonanalytic pair $(z_j, z_j^*)$. The adjoint-phase rule \eqref{eqn:Bloch-adjoint-test-phase} instead defines a Petrov--Galerkin problem: the basis functions carry multiplier $z_j$, whereas the test functions belong to the distinct adjoint Bloch space with multiplier ${z_j^*}^{-1}$. Because $W_+^*=z_j^{-1}W_-^*$, the assembled dependence is then Laurent-polynomial in $z_j$ and $z_j^{-1}$, with no independent $z_j^*$ parameter; the price is a non-Hermitian weak problem.

\paragraph{Bilinear form}
For \eqref{eqn:Bloch-bilinear}, the paired boundary residual is proportional to $z_jW_+-W_-$, so cancellation requires the reciprocal rule $W_+=z_j^{-1}W_-$. This produces analytic Laurent dependence without complex conjugation, both on and off the unit circle. Along each one-parameter Bloch path used here, multiplication by the unknown multiplier removes the negative power and gives a quadratic eigenproblem in $z_j$. The bilinear form is not a Hilbert-space inner product and the resulting algebra is non-Hermitian.

\begin{table}
\centering
\small
\caption{Bloch basis/test pairings for a basis multiplier $z$. The listed test multiplier is carried by $W$ before any conjugation in the form.}
\label{table:Bloch-form-comparison}
\begin{tabular}{|p{0.15\textwidth}|p{0.13\textwidth}|p{0.20\textwidth}|p{0.38\textwidth}|}
\hline
Form & Regime & \mbox{Test multiplier} & Principal consequence \\
\hline
Sesquilinear & $|z|=1$ & ${z^*}^{-1}=z$ & Same approximation and test Bloch space; Hermitian structure for lossless reciprocal media. \\
\hline
Sesquilinear & $|z|\ne1$ & ${z^*}^{-1}$ & Adjoint-phase Petrov--Galerkin problem; Laurent-polynomial dependence after conjugation, but non-Hermitian. \\
\hline
Bilinear & any $z\ne0$ & $z^{-1}$ & One reciprocal rule on and off the unit circle; analytic Laurent dependence, but no Hermitian energy interpretation. \\
\hline
\end{tabular}
\end{table}

Difference schemes such as FLAME and GEFLAME do not utilize weak forms and hence are not hindered by this complex conjugation. Still, variational formulations have their advantages: they are more amenable to stability and error analysis and, depending on the case under consideration, are more likely to be structure-preserving. 

The Trefftz-DG simulation below uses the bilinear formulation to obtain a polynomial pencil with analytic dependence on the Bloch multiplier both on and off the unit circle.
%
\subsubsection{Conforming variational Schur/DtN reduction}\label{sec:conforming-Schur-DtN}
%
Temporarily keeping opposite-side degrees of freedom uncoupled, a conforming discretization yields
\begin{equation}\label{eqn:boundary-interior-block-system}
   \begin{bmatrix}
      A_{bb} & A_{bi}\\
      A_{ib} & A_{ii}
    \end{bmatrix}
   \begin{bmatrix}
      u_b\\ u_i
   \end{bmatrix}
   \,=\, 0.
\end{equation}
For a prescribed boundary vector, $u_i=-A_{ii}^{-1}A_{ib}u_b$, and the retained boundary equation is
\begin{equation}\label{eqn:Bloch-Schur}
    S u_b=0,
    \qquad 
    S \,=\, A_{bb}-A_{bi}A_{ii}^{-1}A_{ib}.
\end{equation}
For a variational discretization, $Su_b$ is the residual in the retained boundary test equations after the interior equations have been solved. In this sense, $S$ is a finite-dimensional analog of the Steklov--Poincar\'e or Dirichlet-to-Neumann operator: its entries are boundary functionals, not pointwise normal derivatives \cite{QuarteroniValli1991,QuarteroniValli1999,Yuan-DtN-BandGaps-2006,Yuan-DtN-2007,Yuan-PhC-inteqn-DtN-2008}.

The same elimination applied to FLAME or another difference scheme still produces a discrete DtN-type map: prescribed boundary samples are extended by solving the interior difference equations, and the Schur complement returns the boundary-equation residuals. Thus both constructions are discrete boundary response operators, but only the variational one is directly a weak DtN map.

A related Schur-complement formulation was previously applied to photonic-crystal waveguides \cite{KlindworthSchmidtFliss2014}. There the primary unknown is the lattice-periodic Bloch factor; the  wavenumber $q$ enters the volume terms, and consequently Schur reduction retains the undesirable nonlinear dependence on $q$ in the eigenproblem. Also, periodic-factor \( q(\omega) \) discretizations may exhibit phantom-mode artifacts \cite{TsukermanFullFieldPeriodic2026}. In the full-field formulation used here, $q$ enters only through periodic-boundary coupling, and Schur reduction preserves the polynomial eigenproblem in \( z = \exp(\ii qa) \), avoiding the periodic-factor phantom-mode mechanism \cite{TsukermanFullFieldPeriodic2026}.

In the full-field problem, the Bloch multiplier enters only through periodic-boundary coupling, and therefore the leading and trailing coefficient matrices of the unreduced quadratic pencil have zero interior blocks. An additional benefit of Schur reduction in this case is elimination of this zero-block structure and the associated infinite eigenvalues, while preserving the finite physical roots. 
%
\FloatBarrier
\subsection{A 2D Trefftz-DG Bloch example}\label{sec:two-bar-dg-Bloch}
%
The 2D example in this section is a proof of concept for the algorithmic combination of (i) the bilinear Bloch formulation in \sectref{sec:Sesquilinear-vs-bilinear}, (ii) the Trefftz-DG method, and (iii) the Schur-DtN reduction. 

\begin{figure}[H]
\centering
\includegraphics[width=0.48\linewidth]{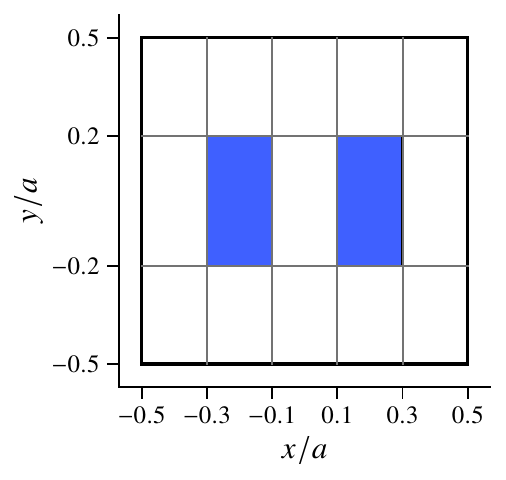}
\caption{Two-bar Trefftz-DG Bloch example: unit cell and material-aligned $5\times3$ level-0 partition. The two central blue rectangles are the dielectric bars.}
\label{fig:two-bar-cell-partition}
\end{figure}

\paragraph{Geometry and local approximation}
The unit cell contains two rectangular dielectric bars and is partitioned by a material-aligned rectangular grid. In the computational domain $\Omega = [-1/2,1/2]^2$, the bars have the relative permittivity $\eps = 6$ and are located at
\begin{equation}\label{eqn:two-bar-inclusions}
[-0.3,-0.1]\times[-0.2,0.2],
\quad
[0.1,0.3]\times[-0.2,0.2],
\end{equation}
in an otherwise empty cell. The level-0 partition is the coarse $5 \times 3$ grid in \figref{fig:two-bar-cell-partition}; level~1 is the corresponding  $10 \times 6$ refinement.

Over each homogeneous rectangle $K$, the field is expanded in local plane waves,
\begin{equation}\label{eqn:two-bar-dg-local-space}
  E_h|_K(\bfr) \,=\,
  \sum_{m=1}^{N_{\rm PW}}c_{K,m}
  \exp \left(\ii k_K\bfd_m\cdot(\bfr-\bfr_K)\right),
  \quad k_K=k_0\sqrt{\eps_K}.
\end{equation}
Standard Trefftz-DG and plane-wave DG formulations are normally sesquilinear \cite{HiptmairMoiolaPerugia2011,HiptmairMoiolaPerugia2016}. For real wave vectors, $|z|=1$, so the same Bloch phase can be used for the basis and test functions and the usual sesquilinear Galerkin construction is adequate \cite{LuCesmeliogluVanderVegtXu2017}. As already noted, for a genuinely complex wave vector off the unit circle, however, that same-space construction introduces both $z$ and $z^*$ and does not cancel the periodic-face residual. If $q$ is purely imaginary, this is a restricted exception in which $z$ is real and conjugation creates no independent parameter, but reciprocal phases for the basis and test functions are still required. To make the Trefftz-DG simulation applicable to general complex bands, the reciprocal-phase bilinear formulation of \sectref{sec:Sesquilinear-vs-bilinear} is used here.

The bilinear skeleton form for Trefftz-DG is
\begin{equation}\label{eqn:two-bar-dg-skeleton-structural}
   b_h(u,v) \,=\, \sum_{F\in\calF_h}
   \int_F \bigg(
   \{\partial_n u\}[v]+\{\partial_n v\}[u]
   +\alpha k_F[u][v]
   +\frac{\beta}{k_F}[\partial_n u][\partial_n v]
   \bigg) \, \dd s
\end{equation}
where for brevity the periodic pairing of faces is absorbed into the jump and average notation. To avoid a proliferation of cases and parameters, fixed penalties $\alpha = \beta = 0.5$ are used in all simulations.

\begin{figure}
\centering
\includegraphics[width=0.8\linewidth]{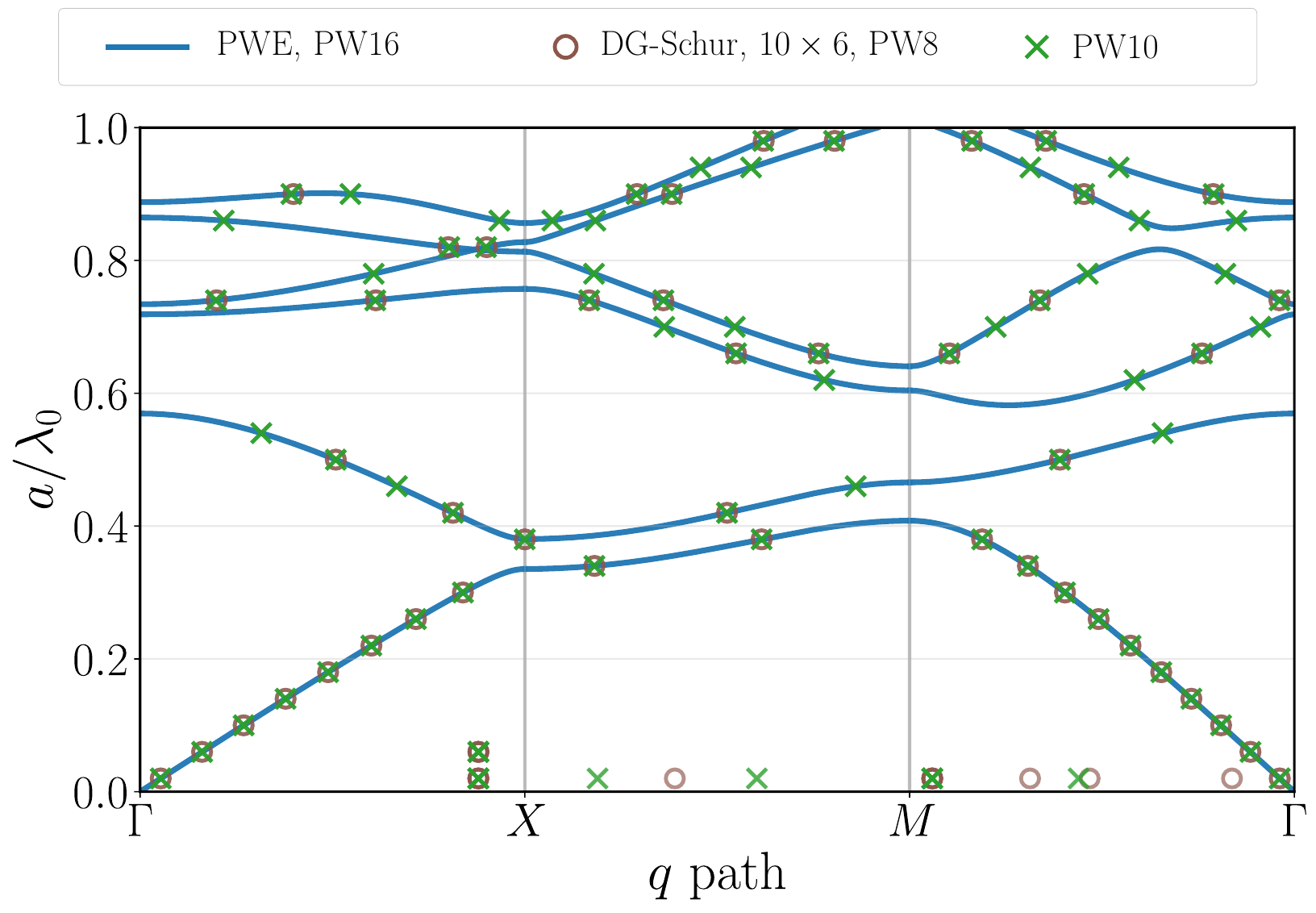}
\caption{Two-bar Trefftz-DG Bloch example: $10\times6$ material-aligned partition (60 rectangular elements). Solid curves are the PWE PW16 comparison curves; markers are DG roots retained after the filter $|\operatorname{Im}(qa)|\le0.05$ for PW8 and PW10 local plane-wave bases. The isolated low-frequency markers are numerical outliers discussed in the text.}
\label{fig:two-bar-dg-l1-fullpath}
\end{figure}

\paragraph{Reciprocal-multiplier Schur reduction}
With $z_x = \exp(\ii q_xa)$ and $z_y = \exp(\ii q_ya)$, the assembled matrix has the Laurent form
\begin{equation}\label{eqn:two-bar-dg-laurent}
A(z_x,z_y)=A_0+z_xA_{x,+}+z_x^{-1}A_{x,-}
+z_yA_{y,+}+z_y^{-1}A_{y,-}.
\end{equation}
Along the three path segments this gives the quadratic pencils
\begin{align}
   \Gamma X: &
          \quad z^2A_{x,+} + z(A_0+A_{y,+} + A_{y,-}) + A_{x,-},\\
   XM: &  \quad z^2A_{y,+} + z(A_0-A_{x,+} - A_{x,-}) + A_{y,-},\\
   M\Gamma: & \quad z^2(A_{x,+} + A_{y,+}) + zA_0+(A_{x,-} + A_{y,-}).
\end{align}
After partitioning the plane-wave coefficients into boundary-layer and eliminated interior sets, write a segment pencil as $Q(z)=z^2P+zB+M$. All multiplier-dependent $P$ and $M$ blocks are supported on the  periodic-boundary layer, and the interior blocks are zero:
\begin{equation}\label{eqn:multiplier-dependent-block-support}
    P_{ii} = M_{ii} = P_{ib} = P_{bi} = M_{ib} = M_{bi} = 0.
\end{equation}
Without Schur elimination, these zero interior blocks generate spurious infinite or NaN eigenvalues. Schur reduction removes those blocks:
\begin{equation}\label{eqn:two-bar-dg-Schur}
   \widehat Q(z) \,=\,
   P_{bb} z^2 \,+\, \left(B_{bb} - B_{bi}B_{ii}^{-1}B_{ib} \right) z + M_{bb}.
\end{equation}
%
\paragraph{Numerical results}
%
The Bloch bands for the $10\times6$ material-aligned partition (60 rectangular elements) are shown in \figref{fig:two-bar-dg-l1-fullpath}. A plane-wave-expansion (PWE) calculation is included as an independent comparison.
PW16 denotes the reciprocal set $m, n = -16, \ldots, 16$ with $33 \times 33 = 1,089$ plane waves. PW20 and PW24 analogously use 1,681 and 2,401 plane waves. For the first six plotted bands, increasing the cutoff from PW16 to PW20 changes $a/\lambda_0$ by at most $2.85 \times 10^{-5}$, with RMS movement $1.37 \times 10^{-5}$; the PW20-to-PW24 movements are $1.36 \times 10^{-5}$ and $6.23 \times 10^{-6}$. These changes are immaterial at the scale of \figref{fig:two-bar-dg-l1-fullpath}. The principal DG branches and the cutoff-stable PWE curves are in good agreement.

A few isolated DG roots remain near zero frequency, chiefly on the $XM$ and $M \Gamma$ segments. These are numerical outliers, plausibly associated with low-frequency degeneration of the propagating-plane-wave Trefftz basis. These nonphysical data points can be removed by continuity-based branch tracking, but are retained in the diagram for the sake of numerical completeness.

%
\FloatBarrier
\section{Conclusions}\label{sec:Conclusions}
%
\subsection{Main findings and contributions}
The paper demonstrates that Trefftz and quasi-Trefftz approximations can be embedded in markedly different discretization mechanisms: nonconforming difference schemes, conforming trace-lift elements, weak trace-matching elements, and broken discontinuous Galerkin spaces. Established FLAME, $T$-element, Trefftz-DG, and Schur--DtN constructions provide the background; the new developments are GEFLAME, TFF, and, for periodic media, the specific integration of the full-field $q(\omega)$ formulation, reciprocal-phase Trefftz coupling, and Schur--DtN reduction. Although the examples are electromagnetic, they extend to analogous linear problems in other areas of physics and engineering.

GEFLAME treats the field and its first derivatives as co-located unknowns. In the single- and four-cylinder tests, it produces field and gradient errors several orders of magnitude below those of the particular quadratic and trace-based conforming discretizations tested at comparable algebraic cost. The node--edge and Cartesian variants further show that the choice and placement of degrees of freedom and sampling geometry are consequential.

TFF supplies a conforming quasi-Trefftz construction on mixed polygonal and interface-cut meshes by lifting prescribed traces through local FLAME solves. The localized reentrant-corner experiment shows that a small patch of problem-adapted singular enrichment can remove the dominant error. For periodic media, the specific new combination is a full-field $q(\omega)$ Bloch formulation, reciprocal-phase bilinear Trefftz-DG coupling, and Schur reduction to an assembled boundary-response map. Besides reducing the dimension of the polynomial eigensystem, the Schur step removes the zero interior coefficient blocks that otherwise generate spurious infinite or NaN eigensolutions.
%
\subsection{Conformity and the trace bottleneck}
\label{sec:trace-bottleneck}
%
The numerical results indicate that constraining Trefftz approximations, strongly or weakly, to prescribed polynomial edge traces can make trace approximation the dominant error and prevent the element-interior Trefftz space from realizing its native high-order or exponential convergence. Conformity remains valuable for analysis and structure-preserving mechanisms, but is subject to this approximation bottleneck.

This conclusion applies to the $T$ and TFF elements tested here, but not directly to Trefftz-DG, which makes a different tradeoff. Trefftz-DG works in broken elementwise Trefftz spaces and avoids a fixed polynomial trace; instead, it must control field and flux jumps, penalties and conditioning. 
%
\subsection{Limitations and future work}
%
The paper focuses deliberately on Trefftz approximations and does not explore the established enhancement machinery of classical FEM, including isoparametric geometry, $hp$-adaptivity, and basis enrichment. Given the many factors affecting computational cost and accuracy, no general ranking of the methods is claimed. Classical polynomial FEM remains the practical workhorse in many engineering and physics applications, whether with $hp$-refinement or with quadratic elements as a simple compromise. That said, the orders-of-magnitude error reductions demonstrated by GEFLAME in the scattering examples identify a potentially useful niche for the method.

Extension of these methods to three-dimensional problems is an important direction for future work. The key ideas do carry over from 2D to 3D, but significant hurdles remain: system size and matrix density, the cost of direct solvers, convergence of iterative solvers for wave problems, and derivation of suitable Trefftz bases. Plane waves, both traveling and evanescent, remain relatively simple in homogeneous regions, whereas three-dimensional solutions near interfaces, such as Mie-type modes, are substantially more cumbersome.

These difficulties notwithstanding, the cost of local approximation and matrix assembly is asymptotically much lower than that of the global solve. This makes further exploration of GEFLAME and related Trefftz-approximation methods in 3D electrodynamics reasonable, probably with edge-based rather than nodal-based degrees of freedom.

\bibliographystyle{elsarticle-num}
\bibliography{Trefftz_methods}

\end{document}